\documentclass[%
reprint,
superscriptaddress,
amsmath,amssymb,
aps,
]{revtex4-1}

\usepackage{graphicx}
\usepackage{dcolumn}
\usepackage{bm}
\usepackage{hyperref}

\begin{document}

\preprint{APS/123-QED}

\title{Crossing integer spin resonance at VEPP-4M with conservation of beam polarization
} 

\newcommand{\BINP}{Budker Institute of Nuclear Physics, 11, Lavrentiev prospect, Novosibirsk, 630090, Russia}
\newcommand{\NSU}{Novosibirsk State University, 2, Pirogova street, Novosibirsk, 630090, Russia}
\newcommand{\NSTU}{Novosibirsk State Technical University, 20, Karl Marx prospect, Novosibirsk, 630092, Russia}
\author{A.K.~Barladyan}
\affiliation{\BINP}
\author{A.Yu.~Barnyakov}
\affiliation{\BINP}
\affiliation{\NSU}
\author{V.E.~Blinov}
\affiliation{\BINP}
\affiliation{\NSU}
\affiliation{\NSTU}
\author{S.A.~Glukhov}
\affiliation{\BINP}
\affiliation{\BINP}
\author{S.E.~Karnaev}
\affiliation{\BINP}
\author{E.B.~Levichev}
\affiliation{\BINP}
\affiliation{\NSTU}
\author{S.A.~Nikitin}
\email[Email:]{S.A.Nikitins@inp.nsk.su},
\affiliation{\BINP}
\author{I.B.~Nikolaev}
\email[Email:]{I.B.Nikolaev@inp.nsk.su},
\affiliation{\BINP}
\affiliation{\NSU}
\author{I.N.~Okunev}
\affiliation{\BINP}
\author{P.A.~Piminov}
\affiliation{\BINP}
\author{A.G.~Shamov}
\affiliation{\BINP}
\affiliation{\NSU}
\author{A.N.Zhuravlev}
\affiliation{\BINP}
\affiliation{\NSU}

\date{\today}

\begin{abstract}
A method proposed  to preserve the electron beam polarization at the VEPP-4M
collider during acceleration with crossing the integer (imperfection) spin
resonance at energy E=1763 MeV has been successfully applied. It is based on
full decompensation of the $ 0.6\times3.3$ Tesla$\times$meter integral of the
KEDR detector longitudinal magnetic field due to the anti-solenoids
'switched-off'.
\begin{description}
\item[PACS numbers]
{
  29.20.D-,  
  29.20.db,  
  29.27.Fh,  
  29.27.Hj   
}
\end{description}
\end{abstract}

\maketitle


\section{\label{sec:Motivation}Motivation}
\begin{figure}[htb]
\centering
\includegraphics[width=80mm]{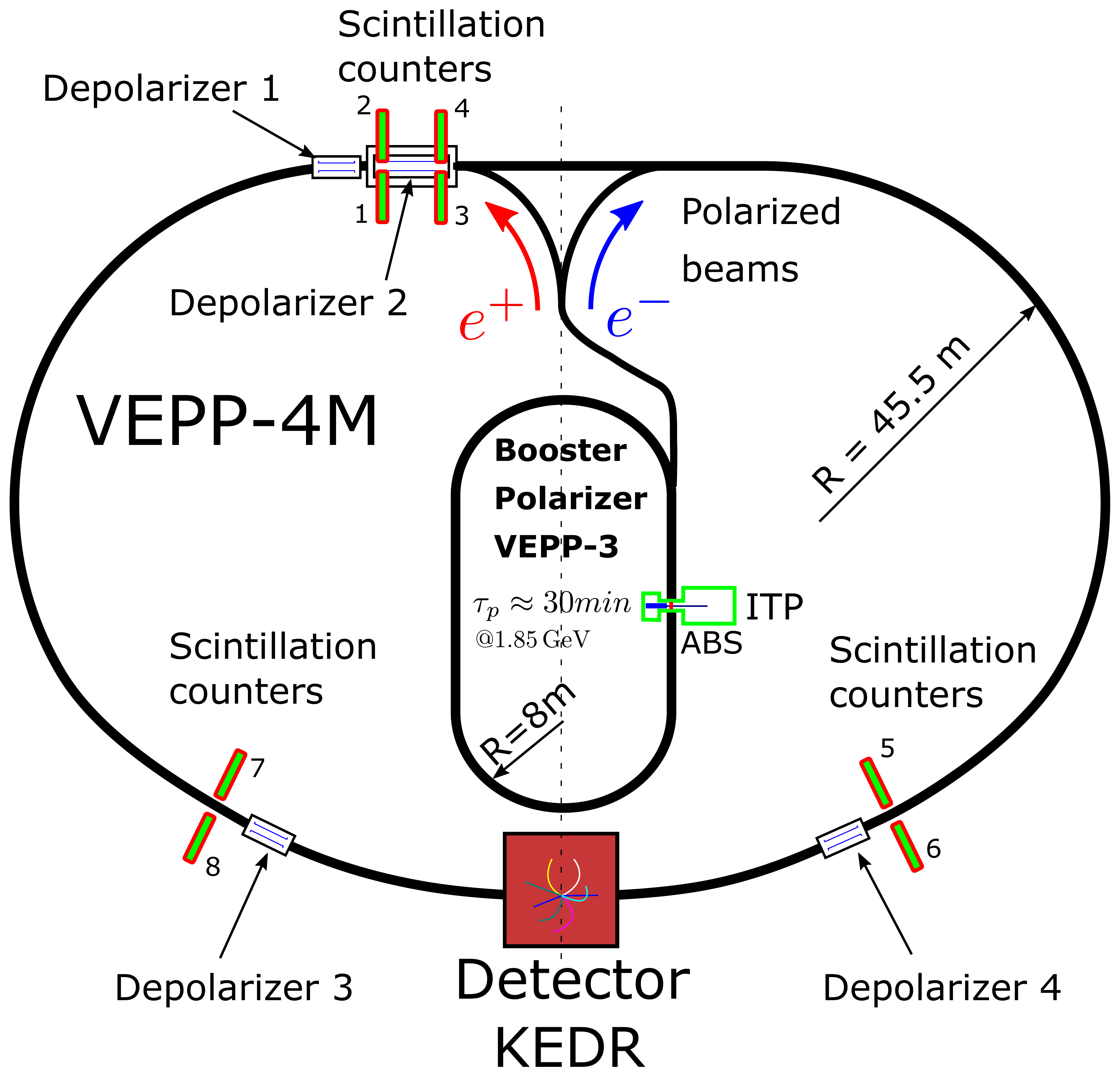}
\caption{
Scheme of the VEPP-4 complex from the point of view of polarization
experiments. ITP is the internal target setup based on the use of a jet of
polarized deuterium atoms  from the Atomic Beam Source (ABS). } 
\label{VEPP-4}
\end{figure}

\begin{figure}[htb]
\centering
\includegraphics[width=75mm]{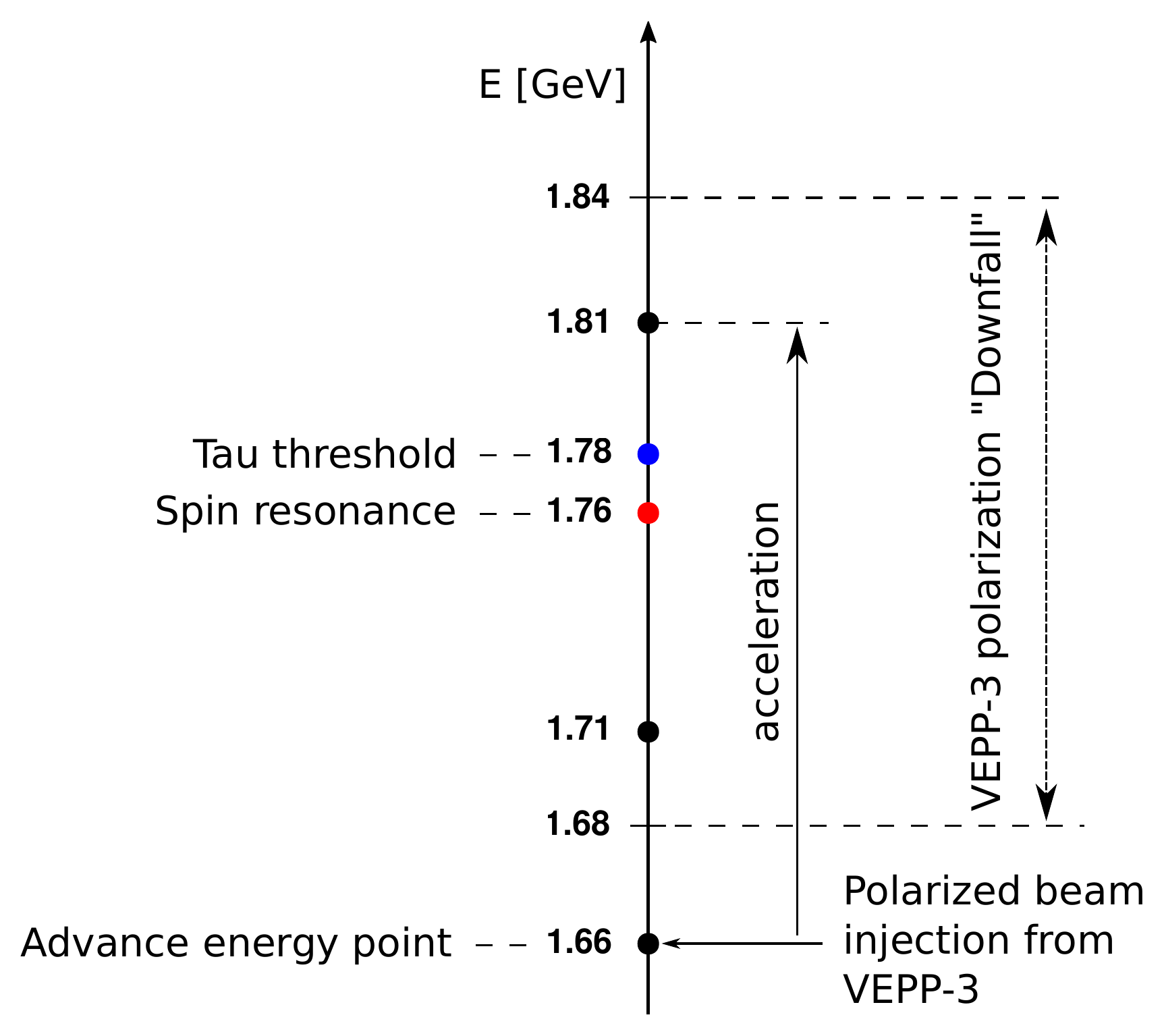}
\caption{Diagram of energy points in the experiment.}
\label{fig:Diag}
\end{figure}

The set of the beam energy values in the Hadron-Muon Branching Ratio measurement
with the KEDR detector~\cite{KEDR} at the electron-positron VEPP-M collider~\cite{VEPP-4M} 
in the region between $J/\psi$ and $\psi '$ resonances includes
several critical points, in particular,  $E=1764$~MeV and $1814$~MeV. The
beam energy calibration in this experiment is performed using the Resonant
Depolarization technique (RD), and thus polarized beams are required. Polarization is
obtained due to the natural radiation mechanism at the VEPP-3 booster storage
ring (see Fig.~\ref{VEPP-4}).  Both mentioned energy values are in the so-called
'Polarization Downfall' (Fig.~\ref{fig:Diag}), which is the VEPP-3 energy
range of approximately 160 MeV width,  where  obtaining of a fairly high degree of
polarization is significantly hampered because of the strong depolarization
effect of different field imperfections. The 'Polarization Downfall' range was
found  in the 2003 year experiment  with the polarimeter based on an internal
polarized target~\cite{DYUG}.

The center of that critical range is the energy value $E_4=1763$~MeV, which
corresponds to the forth imperfection spin resonance  $\nu=\nu_k=4$. In a
conventional storage ring without any imperfections, $\nu=\gamma a$ is the spin
tune parameter equal to the number of the spin vector precessions about the
vertical guide field per a turn minus one; $\gamma$ is the Lorentz
factor; $a=(g-2)/2$ is the magnetic dipole moment anomaly.  Nevertheless, one
can obtain polarized beams at VEPP-4M  with the energies from the `Polarization
Downfall', except for a small island in the vicinity of $E_4$ using the
method of `advance energy point'. In the given experiment, the magnetization reversal
cycle of the collider is of the `upper' type. It means that the 'advance energy
points'  should be below  the energies of experiment, as well as below 1660 MeV
because of the `Polarization Downfall' region lower boundary. In the
cases when the method is valid, the beam polarization in VEPP-3 is achieved at
the 'advance' energy. Then the beam is injected into the collider ring. After
that its energy is raised to the energy of the experiment. The radiative spin
relaxation time in the collider ring is two orders larger than that in the
booster ring ($\tau_p\approx 80$~h at $E=1.8$~GeV). This allows us to apply the
RD technique even at rather small detuning from the dangerous spin
resonances, despite the sharp increase in the depolarizing effect of quantum
fluctuations in the presence of guide field imperfections.  For instance, it
was so in the tau-lepton mass measurement experiment~\cite{TAU}  at energies
close to the tau production threshold ($E=1777$ MeV). The RD calibrations were
carried out at the detuning $\delta \nu\approx 0.03$  ($\Delta E\approx 13$ MeV in units of energy) and less from the resonance
$\nu_k=4$. 
The magnetization reversal cycle of the `lower' type was used, and the 'advance' energy was 1.85
GeV.  After the beam was injected into VEPP-4M, its energy decreased to the
energy of the tau production threshold.

In the case under consideration, one could apply a similar method at energy points below 1763 MeV.
However, there was a need for special measures in relation to the energy
$E=1814$ MeV, as well as to the points of energy  somewhat below the $\psi '$
peak  (for example, $E=1839$ MeV). The reason was the necessity to cross the
integer spin resonance at 1763 MeV during acceleration starting from the
'advance' energy.  

\section{\label{sec:Issues} Issues of fast/slow crossing} 

A sufficiently high rate of beam acceleration or decceleration  in a storage
ring can save the polarization of particles  to a considerable extent in
crossing of any spin resonance $\nu_0=\nu_k$.  Generally, an actual  spin tune
$\nu_0$  differs from the parameter $\nu$ defined for storage rings with an
unidirectional guide field. For fast crossing, the following condition must be
fulfilled~\cite{Froissart:1960zz,DERB}: 
\begin{equation}
\frac{d \varepsilon}{dt}=\dot \varepsilon \gg |w_k|^2\omega_0,
\label{eq:eq1}
\end{equation}
where $\varepsilon(t)=|\nu_0(t)-\nu_k|$ is the time-dependent resonant
detuning; $w_k$ is  the resonant harmonic amplitude of the field perturbations;
$\omega_0=2\pi f_0$  is the angular frequency of particle revolution (at
VEPP-4M, $f_0=819$~kHz).  The corresponding loss in the degree of polarization
at a single crossing of the resonance is $\sqrt{\pi |w_k|^2|\omega_0/\dot
\varepsilon}\ll 1$.  The data on the polarization lifetime $\tau_d$ due to
radiative depolarization in the vicinity of $E=1777$ MeV, obtained during the
preparation of the tau-lepton mass measurement  experiment, contain the
information about the natural strength of the spin resonance $\nu_k=\nu=4$.
The polarization lifetime was adjusted to the level $\tau_d\ge
1$~h~\cite{TAU-POL-EPAC-2004}.  One can associate this quantity with the formal
estimate of the resonant spin harmonic amplitude using the known
equation~\cite{DERB2,Derbenev:1973ia}:
\begin{equation}
\tau_d\approx\frac{\tau_p}{1+\frac{11}{18}\frac{|w_k|^2\nu^2}{\varepsilon^4}}
\label{eq:eq2}
\end{equation}
where $\varepsilon=\nu_0-4\approx 0.03 \ll 1$, and $\tau_p$ is the Sokolov-Ternov
polarization time \cite{SokolovTernov}. For VEPP-4M, at $E=1777$ MeV,
$\tau_p=87$ h, $\tau_d=1$ h, and $\varepsilon\approx 0.03$; so the estimate
is  $|w_k|\sim 2.8\times 10^{-3}$.  Therefore, the maximum necessary rate of
the resonance crossing is  $\dot\varepsilon \gg20$\,s$^{-1}$, or $dE/dt \gg
10^{4}$ MeV/s. In practice, the achievable  ramping rate at VEPP-4M does not
exceed 10-20 MeV/s. So, fast crossing of the integer spin resonance $E=1763$~MeV
is impossible. 

If the ramping rate is reduced so that the following condition 
\cite{Froissart:1960zz, DERB, DERB2}
\begin{equation}
\dot\varepsilon \ll |w_k|^2\omega_0, 
\label{eq:eq3}
\end{equation}
is satisfied, then the spin resonance intersection occurs adiabatically slowly.
Basing on the estimates made above, one can conclude that, in principle, a rate of
$1\div10$~MeV/s  may be appropriate.  In the  theoretical limit
of the adiabatic crossing without taking into account  the radiation effects,
the polarization retains its value and changes the sign (the spin flip
mode).

Despite the feasibility of the condition of slow crossing, it is necessary to
bear in mind that there is a lower limit on the rate of the crossing because of the
depolarizing effect of radiation diffusion and damping. The radiation
depolarization time related to the vertical closed orbit distortions declines
very quickly with the detuning from the integer spin resonance:
$\tau_d\propto \varepsilon^{4}$.  For example, this time decreases as 16 times
at  $ E=1770$ MeV as compared with $\tau_d=1$~h at $E=1777$~MeV. In
the case of strong resonance, the spin diffusion also depends on the decrement
$\Lambda$, the parameter of radiation damping. While passing the resonant
region $ |\varepsilon |\sim | w_k | $ at $ \omega_0 |w_k|\gg \Lambda$, due to radiation diffusion and damping,  the
depolarization time reaches the minimum
value of $ \tau_d \sim \Lambda^{- 1} \sim 100$~ms~\cite{KondratenkoDT}.
The estimates given in \cite {Nikitin-BINP-2015-1} allow us to conclude that the
adiabatic crossing of the forth spin resonance in VEPP-4M 
will lead to a notable loss of beam polarization.

\section{\label{sec:SpinKinem}Spin kinematics at KEDR detector field decompensation}

A simple approach to preserve the VEPP-4M beam polarization in the conditions
under consideration was proposed and numerically substantiated
in~\cite{Nikitin-BINP-2015-1} and then successfully implemented in the
experiment~\cite{Barladyan:IPAC-17}.  It is based on the well-known idea of a
partial Siberian snake~\cite{Derb1977}.

If one switches off the current in the anti-solenoid coils, the 0.6 Tesla KEDR
detector longitudinal magnetic field integral becomes uncompensated, which
results in the rotation of the spin around the velocity through  an angle $\varphi$.
In particular,  $\varphi\approx0.34$~rad at $E=1.75$~GeV.
\begin{figure}[htb]
\centering
\includegraphics[width=85mm]{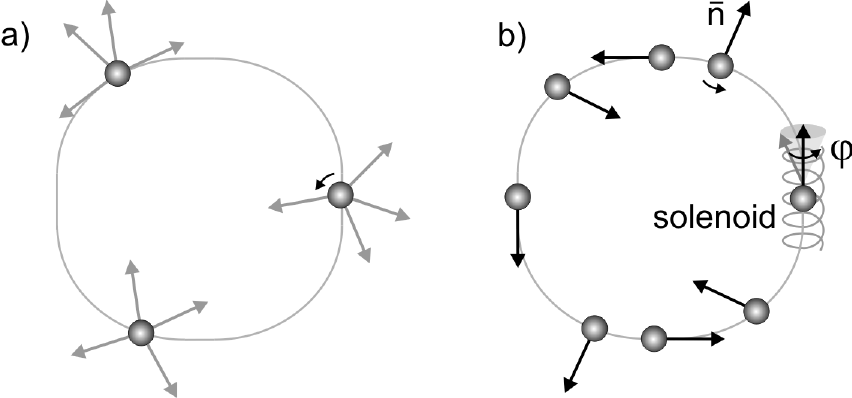}
\caption{
Two cases of the spin kinematics at $\nu=$ integer.  a) There is no a
preferable direction of the spin polarization in an ideal storage ring with a
flat orbit.  b) There is a dynamically stable vector $\vec n$, the periodical
precession axis,  when an arbitrary solenoid is introduced. 
}
\label{fig:Vect}
\end{figure}

Fig.~\ref{fig:Vect} illustrates the principle difference between the cases  of
a storage ring without and with an arbitrary solenoid of $\varphi\neq 0$ at the
condition $\nu=$ integer. We assume that initially there are no field
perturbations in the hypothetical storage ring.  In the first case (a), through
a revolution of a particle , any direction of the polarization vector turns
into itself. By this reason, the spin perturbations growing  due to
fluctuations and the spin tune spread  lead to fast breaking of the beam
polarization. In the second case (b), there exists a general equilibrium
periodical direction of polarization, the $\vec n$ vector, for all beam
particles. This vector rotates in the median plane and is directed along the
particle velocity at the azimuth where the solenoid is located.   Spins with
deviations from $\vec n$, make precession   around it.  Owing to this fact, the
spin motion is stabilized largely. Depolarization can occur through relatively
slow accumulation of small changes in the average projection of the spins  on
the axis $\vec n$.

A longitudinal magnetic field causes a shift of the spin tune with regard to an
unperturbed value $\nu$.  Thus, the non-integer part of the perturbed spin
frequency $\nu_0$ does not take the zero value at the critical point near $E =
1763$ MeV. This fact is used as a basis for preserving the beam polarization
during acceleration. 

\begin{figure}[htb]
\centering
\includegraphics[width=80mm]{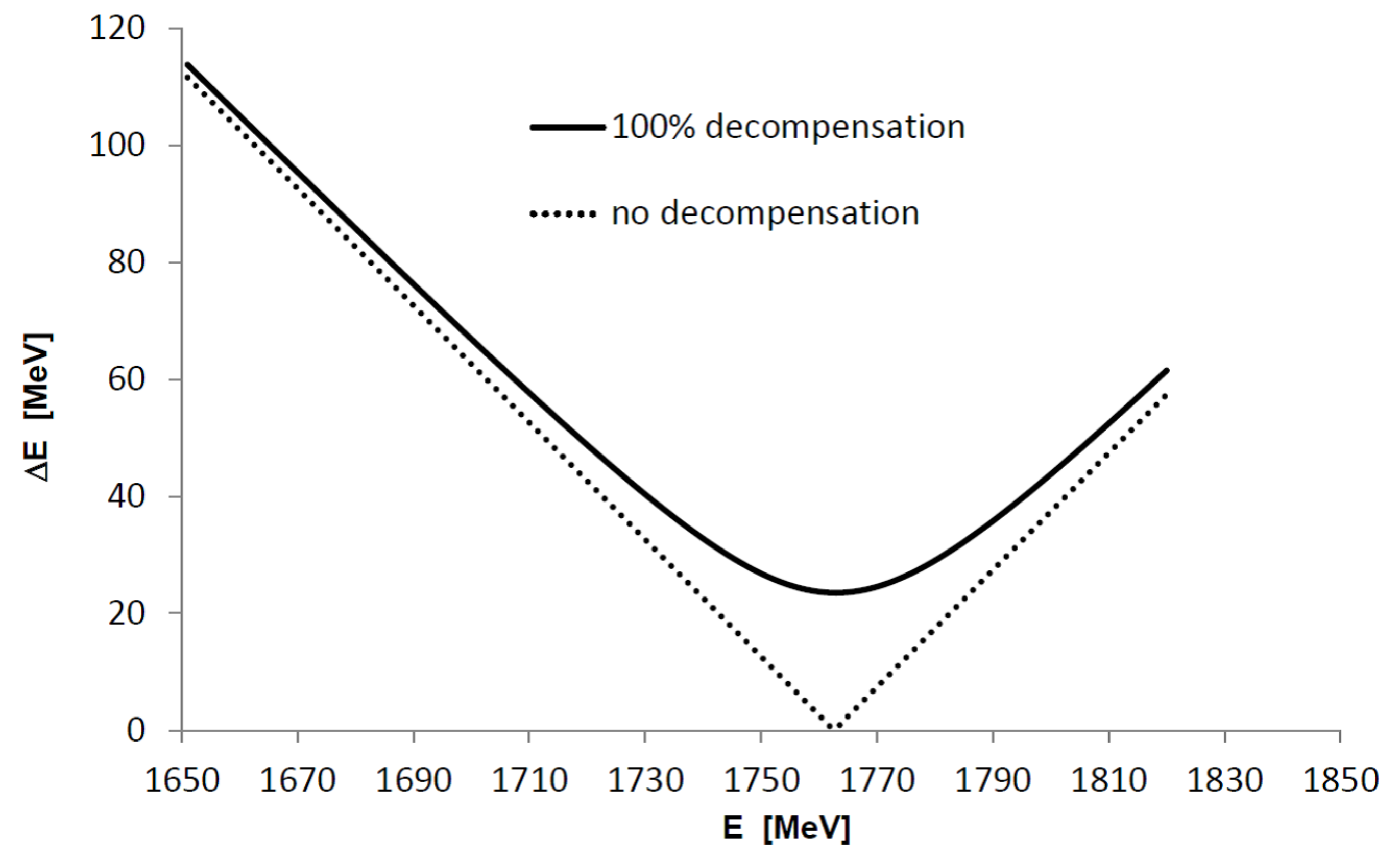}
\caption{Spin tune shift in the energy units vs.\ the beam energy in
  the cases of no and full decompensation of the KEDR detector field integral
  of $0.6\times3.3$~Tesla$\times$meter.
}
\label{fig:Tune}
\end{figure}

The equilibrium polarization axis as a function of the azimuth $\vartheta$ in a
storage ring having at $\vartheta=0$ an insertion with longitudinal magnetic
field  is calculated using the known formulae
\cite{Derbenev:1978hv,NikSaldin}:
\begin{equation}\label{eq:n}
\begin{split}
			n_x(\vartheta)&=\pm\frac{\sin{\nu(\vartheta-\pi)}}{\sin{}\xi}\cdot\sin{\frac{\varphi}{2}}, \\  
			n_y(\vartheta)&=\mp\frac{\cos{\nu(\vartheta-\pi)}}{\sin{}\xi}\cdot\sin{\frac{\varphi}{2}}, \\
			n_z(\vartheta)&=\mp\frac{\sin{\pi\nu}}{\sin{}\xi}\cdot\cos{\frac{\varphi}{2}} ,\\
			\sin\xi&=\sqrt{1-\cos^2{\pi\nu}\cos^2{\frac{\varphi}{2}}}.
\end{split}
\end{equation}
Here 
\begin{equation*} 
\varphi\approx \frac{\pi}{4.6\nu}\cdot\int H_{||}ds
\end{equation*} 
is the angle of electron spin rotation in the longitudinal magnetic field with
the integral of $\int H_{||}ds$ in Tesla$\cdot$meter. The symbols $x,y$, and
$z$ denote the horizontal, longitudinal, and vertical orts of the movable
coordinate basis,  respectively.  We use an approximation of an isomagnetic
storage ring in which the azimuth $\vartheta$  equals the angle of the particle
velocity rotation.  The effective spin precession  tune is determined from the
equation
\begin{equation*}
\cos{\pi\nu_0}=\cos{\pi\nu}\cos{\frac{\varphi}{2}}. 
\end{equation*} 
In our case, $\int H_{||}ds=H_{KEDR}\cdot L_{eff}$, where $H_{KEDR}=0.6$~Tesla,
the KEDR detector field; the effective KEDR solenoid length $L_{eff}=3.3$~m  if
the anti-solenoids are switched off. 
Fig.~\ref{fig:Tune} shows the calculated spin tune shift $|\nu_0-\nu|\propto
\Delta E$ versus the beam energy at full decompensation.  This shift  is about
$\Delta E=22$~MeV in the vicinity of the critical energy 1763~MeV.

\section{\label{sec:Combination}Combination spin resonances}

In our experiment, there is  an intersection of some combination spin
resonances $\nu_0\approx\nu=\nu_k$, which depend on the betatron tunes
$\nu_x=8.536$ and $\nu_z=7.572$.  Because of the relatively narrow range of
energy adjustment,  the main spin-betatron resonances $\nu\pm\nu_z=k_z$  and
$\nu\pm\nu_{x}=k_x$ did not fall into it. According to the results of the
experiment, it can be argued that the subsequent weaker resonances
$\nu+\nu_x+\nu_z=20$ (1715~MeV) and $\nu-\nu_x-\nu_z=-12$ (1810.2~MeV) were
successfully crossed. To comment on this, we estimate the amplitude of the
harmonic of the spin resonances of the $\nu\pm\nu_x\pm\nu_z=k_{xz}$ type by the
formula
\begin{equation*}\label{eq:harm}
|w_{k_{xz}}|\sim |\langle \nu  h\sqrt{\sigma_x\sigma_z} F^\nu e^{i[\pm (\mu_z+\mu_x)\mp (\nu_x+\nu_z\pm k_{xz})\vartheta]}\rangle|\sim 10^{-7}.
\end{equation*}
Here, $h=\partial ^2 H_z/\partial x^2$ is the quadratic non-linearity due to
the sextupole correction (in the units of the mean field $\langle H_z\rangle$);
$\sigma_x$ and $\sigma_z$ are the transverse beam sizes; $\mu_{x,z}$ are the
betatron phase advances; $\langle...\rangle$  is averaging over azimuth; $F^\nu
(\vartheta)$ is the periodic spin response function~\cite{DERB2}.  The function
$F_\nu$ takes into account the depolarization effect of vertical betatron
oscillations excited by any local perturbation.  The results of the calculation
of this characteristic for VEPP-4M are given, for example, in
\cite{MRD_NPPP_2016}. Fast crossing of such weak resonances  becomes possible,
starting from the very low rates of change in the energy: $dE/dt \gg 10$~eV/s.

\section{\label{sec:Radiative}Radiative depolarization rate during acceleration}

The characteristic time $\tau_d$ of the radiative depolarization due to
quantum fluctuations in the presence of a strong perturbation in the form of the
KEDR detector longitudinal field can be found from the generalized equation
\cite{Derbenev:1973ia}: 
\begin{equation}\label{DKformK}
\tau_d\approx\frac{\tau_p}{\left< 1-\frac{2}{9}(\vec n \vec \beta)^2+\frac{11}{18}\vec d^2 \right>}.
\end{equation}
Here $\tau_p$ is the Sokolov-Ternov polarization time (it is proportional to
$E^{-5}$ and amounts to 72~h at the VEPP-4M energy of 1.85~GeV);  $\vec d^2$ is
the square of  the periodical  spin-orbit coupling vector function of the
azimuth. In our case, the spin-orbit coupling is excited by the uncompensated
part of the  KEDR field integral. It can be found as the derivative of
vector~(\ref{eq:n}) with respect to the Lorentz-factor $\gamma$ of
particles~\cite{NikSaldin}: 
	\begin{equation}\label{dndg}
		\begin{split}
			d_x&=\gamma\frac{\partial n_x}{\partial \gamma}=\pm\bigg\{F\sin{\nu(\vartheta-\pi)}\cdot\sin{\frac{\varphi}{2}}+\\
			& \quad  \frac{1}{\sin{\xi}}\left[\nu(\vartheta-\pi)\cdot\cos{\nu(\vartheta-\pi)}\cdot\sin{\frac{\varphi}{2}}-\right.\\
	 & \quad \left.\frac{\varphi}{2}\cdot\sin{\nu(\vartheta-\pi)}\cdot\cos{\frac{\varphi}{2}}\right] \bigg\},\\
			d_y&=\gamma\frac{\partial n_y}{\partial \gamma}=\mp\bigg\{F\cos{\nu(\vartheta-\pi)}\cdot\sin{\frac{\varphi}{2}}-\\
			& \quad  \frac{1}{\sin{\xi}}\left[\nu(\vartheta-\pi)\cdot\sin{\nu(\vartheta-\pi)}\cdot\sin{\frac{\varphi}{2}}+\right.\\
	    & \quad \left.\frac{\varphi}{2}\cdot\cos{\nu(\vartheta-\pi)}\cdot\cos{\frac{\varphi}{2}}\right] \bigg\},\\
			d_z & =\gamma\frac{\partial n_z}{\partial \gamma}=\mp\bigg\{F\sin{\pi\nu}\cdot\cos{\frac{\varphi}{2}}+\\
				  & \quad \frac{1}{\sin{\xi}}\left[\pi\nu\cdot\cos{\pi\nu}\cdot\cos{\frac{\varphi}{2}}+\right.\\
	        & \quad \left.\frac{\varphi}{2}\cdot\sin{\pi\nu}\cdot\sin{\frac{\varphi}{2}}\right] \bigg\},\\
			F   & =-\frac{1}{2\sin^3{\xi}}\Big(\pi\nu\sin{2\pi\nu}\cdot\cos^2{\frac{\varphi}{2}}- \\
					 & \quad - \frac{\varphi}{2}\sin{\varphi}\cdot\cos^2{\pi\nu}\Big).
\end{split}
\end{equation}
Formula (\ref{DKformK}) is valid outside of spin resonances. We use it together
with  set of  equations (\ref{dndg}) to calculate the radiative diffusion of
polarization during acceleration, since in the presence of a solenoid, the
effective spin precession frequency does not take integer values.

Another contribution to the spin-orbit coupling is given by the  betatron
oscillations excited by quantum fluctuations.  In consideration of various
cases with spin rotators based on solenoids, the contribution of betatron
oscillations far from the spin-betatron resonances to the depolarization rate
is small as compared with the effect of the dependence of the polarization axis
$\vec{n}$ on the energy~\cite{NikSaldin,NikCtau,NikRupac}. By this reason we
neglect the betatron oscillations.

\begin{figure}[htb]
\centering
\includegraphics[width=85mm]{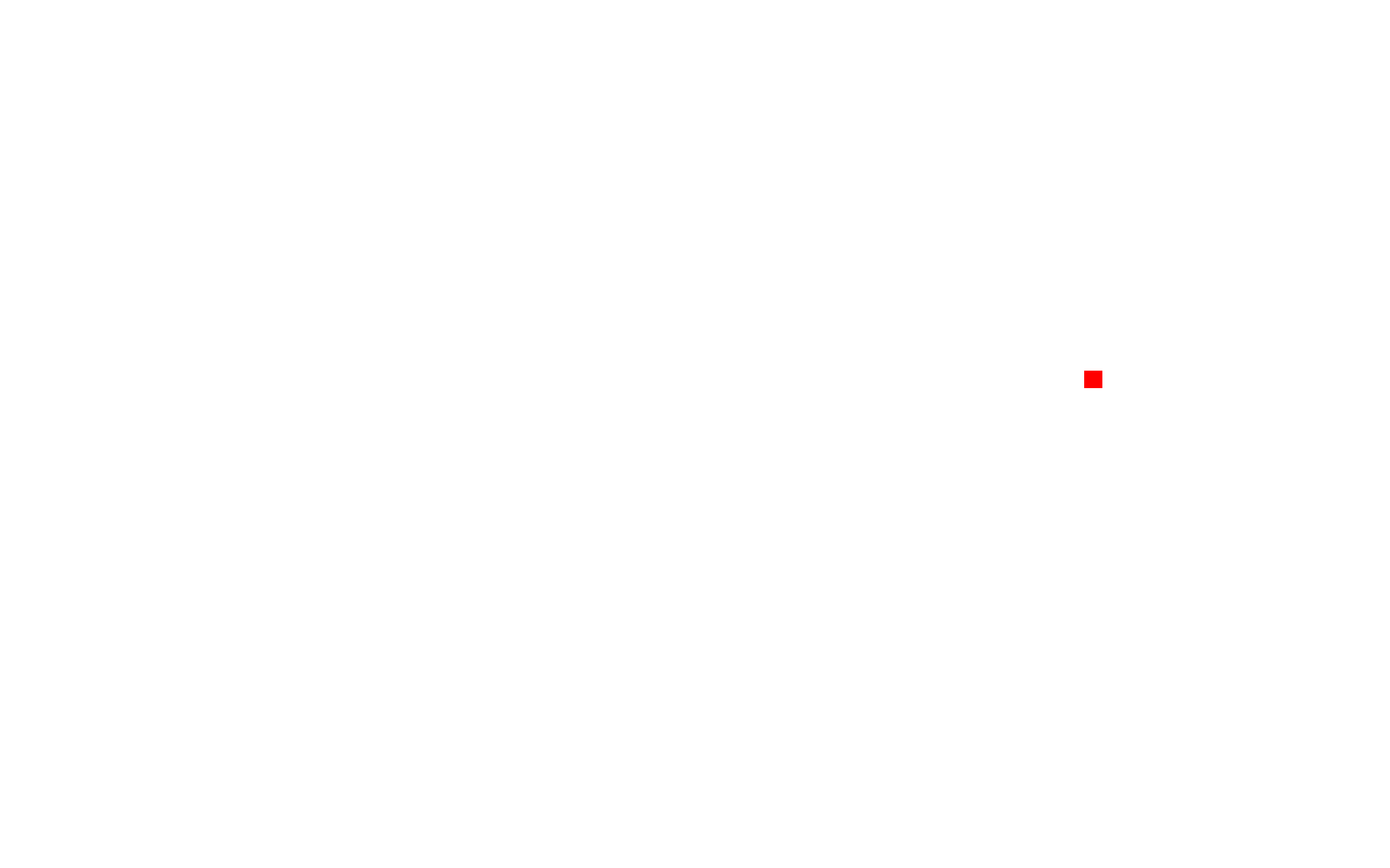}
\caption{The calculated radiative depolarization time vs. the beam energy under
the influence of 0.6~T KEDR field decompensation. The separate point shows the
measured value at an energy of 1806~MeV (estimated value of 1407~s). }
\label{fig:Time}
\end{figure}

The depolarization time is calculated using formulas
(\ref{DKformK},\ref{dndg}) and  plotted in Fig.~\ref{fig:Time} versus the beam
energy at $100\%$  and $50\%$ decompensation of the KEDR field
integral.  The minimum depolarization time $\tau_d = 20$ s. The width of the
energy area where $20\,\mbox{s}<\tau_d<100$\,s is about   30 MeV. It takes about 30 s
 to cross this area at a nominal rate of energy change $dE/dt =1$~MeV/s.

At 1810~MeV, the time $\tau_d$ becomes long enough, 2000~s. This gives a chance
to measure the energy from the spin frequency using the  RD technique after the
end of the acceleration process, followed by the restoration of the field in
the anti-solenoids. 

\begin{figure}[htb]
\centering
\includegraphics[width=80mm]{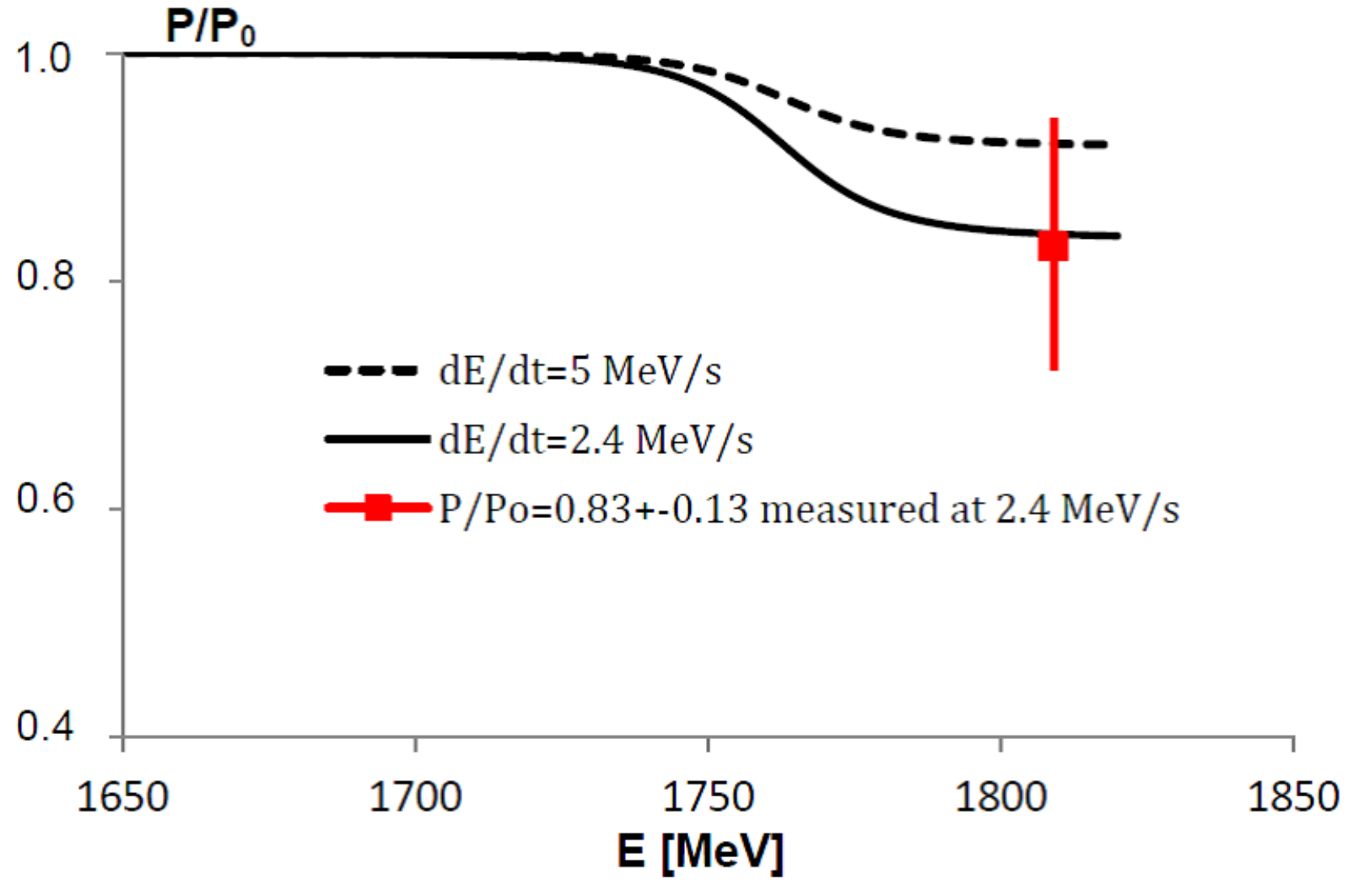}
\caption{The calculated change  in the degree of polarization relative to the
initial one in the course of beam acceleration at a rate characterized by the
parameter $dE/dt$ in the case of complete decompensation of the KEDR field. The
separate point with the error bar presents the generalized  experimental data.}
\label{fig:Degree}
\end{figure}

The theoretical behavior of the polarization degree during acceleration from
the injection ('advance') energy E = 1650~MeV is shown in Fig.~\ref{fig:Degree}
for two values of the acceleration rate.  The current value of the degree in
the units of the initial one ($P_0$) is calculated from the equation
\begin{equation*}
\frac{P}{P_0}\approx \exp{\left[ -\int_{E_1}^{E_2}\frac{dE}{(dE/dt)\cdot\tau_d}\right]}.
\end{equation*}

It is seen that it is advantageous to apply the full decompensation of the KEDR
field and perform acceleration with a rate not below $2$ MeV/s.  In the best
case, it can provide about $80\%$ of the initial polarization degree in the
final state. RD calibration of the beam energy should be performed only after
restoration of the anti-solenoid field, which leads to the elimination of the
spin tuning shift, and with it a systematic error in the energy value..
Moreover, the polarization lifetime increases manifold if the KEDR field is
compensated. 

\section{\label{sec:Compensation}Compensation of betatron coupling from KEDR field}

\begin{figure}[htb]
\centering
\includegraphics[width=85mm]{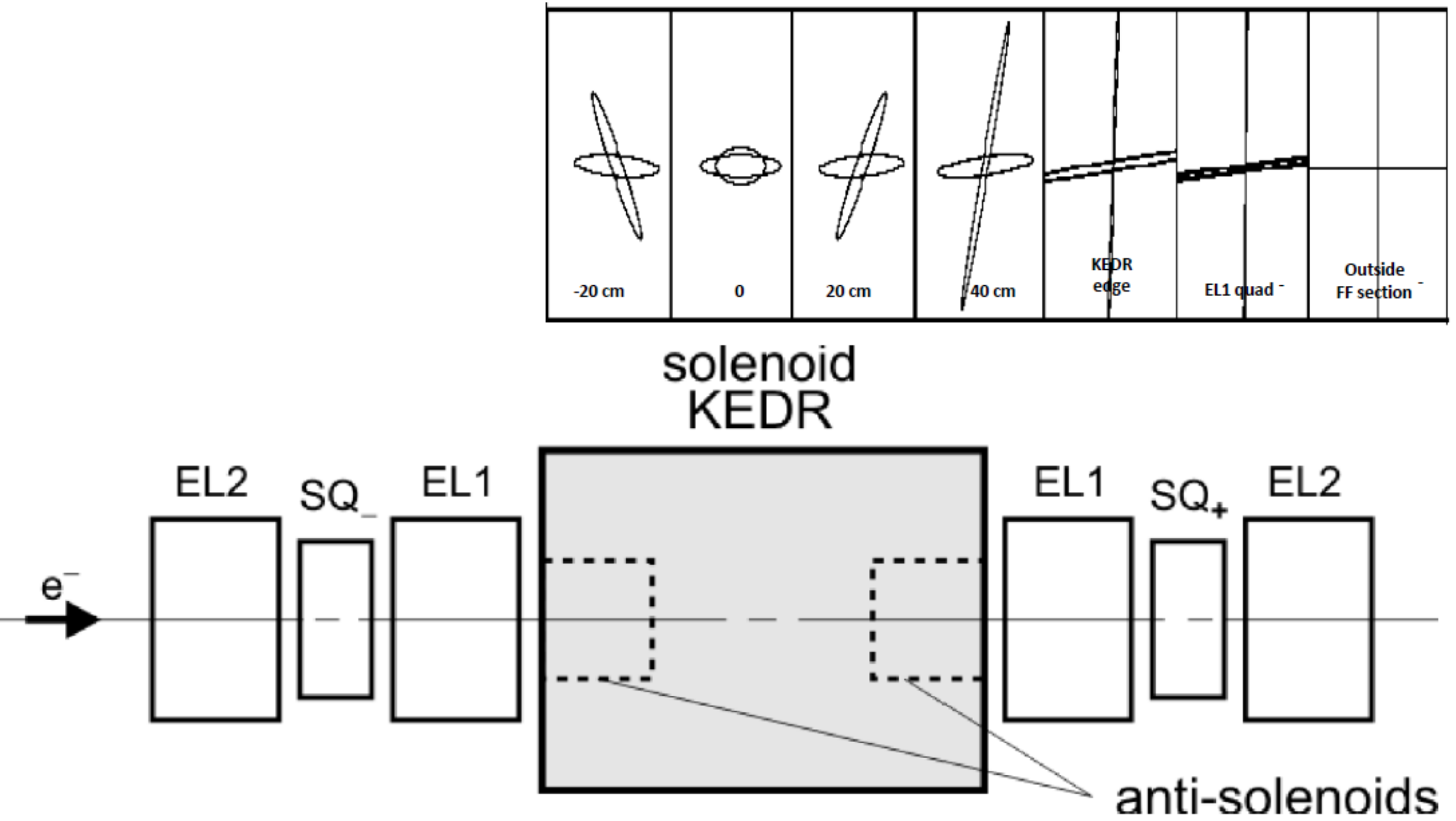}
\caption{The scheme (depicted not to scale) of compensation of the betatron
  coupling caused by  the KEDR main solenoid field  with the help of two skew
  quadrupoles (SQ+ and SQ-) located near the Final Focus lenses. The
  anti-solenoids are  off. Above is a view of  normal betatron  modes in the
  X-Z plane at different distances from the center of the detector.}
\label{fig:Scheme}
\end{figure}

If the anti-solenoids are switched off,  the special measures are needed to
provide the relevant alternative operation modes of VEPP-4M.  It is convenient
to use the scheme of betatron coupling localization by
K.Steffen~\cite{Steffen1982},  which includes two skew quadrupole lenses
(Fig.~\ref{fig:Scheme}).  This scheme has already been successfully tested  at
VEPP-4M~\cite{NikProt1999, Nikitin:2001ej}.

The transport matrix for the vector of betatron variables $(x,x',z,z')$ in the
section from the skew lens SQ$_+$ to SQ$_-$ including KEDR can be approximately
written as
\begin{equation*}
M=Q_-\cdot M_-\cdot M_s\cdot L^2\cdot M_s\cdot M_+\cdot Q_+.
\end{equation*}
Here $Q_\pm$ are the 'thin' skew quad matrices; $L$ is the empty section matrix
for the length $l=L_s/4$;  $L_s=3.3$~m is the effective length of the KEDR main
solenoid ($L_s=2.5$ m  when the anti-solenoids are switched on); $M_s$ is the
half-solenoid matrix in the 'thin magnet' approximation ($\chi=\varphi/2$):
\begin{equation*}
M_s=\left(\begin {array}{cccc}
1 & 0 & - \chi & 0 \\
0 & 1 & 0 & -\chi \\
\chi & 0 & 1 & 0 \\
0 & \chi & 0 & 1
\end{array}\right);
\end{equation*}
$M_\pm$  are the matrices for transformation from the center of the right
(left) half of the KEDR solenoid to the corresponding skew quad.  The skew
quads are placed symmetrically relative to the solenoid in the 'magic'
azimuths, for which some elements of the matrix $M$ exactly or approximately
satisfy a certain equation. The strengths of the $SQ_\pm$ lenses are found from
another equation, proportional to $\chi$, similar in value, and opposite in
sign.  If we set these found skew quad strengths, then the matrix $M$ will not
contain the off-diagonal (coupling) 2x2 blocks or will be close to such kind.
The simplicity of the scheme is based on the mirror symmetry of the magnetic
structure in the section with the solenoid. The betatron coupling is localized
in this section. The vertical and horizontal oscillations excited beyond the
section are mutually independent within the accuracy of the compensation scheme
design and realization. The scheme provides a minimum split of the normal
betatron mode frequencies of the order of $10^{-3}$ (in the units of the
revolution frequency).  If no compensation is applied, this split achieves 0.1,
and this hampers sustainable maintenance of the beam during acceleration. 

\section{\label{sec:Touschek}Touschek polarimeter}

To observe the beam polarization, a  system of absolute calibration of the beam
energy by the frequency of the spin precession was used~\cite{TSH-POL-NIM-2002,
EMS-REVIEW-NIM-2009}. The system includes the Touschek polarimeter, based on
the IBS (Intra-Beam-Scattering) effect, and the TEM wave-based depolarizer. The
polarimeter consists of eight plastic scintillator counters, pushed inside the
accelerator vacuum chamber to registrate the particles scattered from the beam
(Fig.~\ref{VEPP-4}). The rate of counting IBS events depends on the square of
beam polarization~\cite{bayer1969, BKS, STRAKH,TSH-CALC-JETP-2012}. In the
depolarizer, the TEM wave is generated with the help of two parallel,
vertically-spaced conductive plates connected to a variable-frequency RF
generator. We have several such devices located at different azimuths of the
VEPP-4M ring. In various experiments, the most suitable device is used, since
the effectiveness of a transverse field-based depolarizer depends on the energy
and azimuth of its location through the spin response function $F^\nu$. 

Observation of polarization in the described system is possible in two ways.
The first is the resonant depolarization. When the depolarizer frequency is
scanned, at the moment of polarization breaking, a jump occurs in the rate of
counting the Touschek electrons. The magnitude of the depolarization jump is
determined by the square of the polarization degree. To eliminate the influence
of the instability of the beam parameters, we register the ratio of the count
rates from the polarized bunch and unpolarized one. Depending on the scanning
parameters, the accuracy of determination of the particle energy from the
depolarizer  frequency, associated with the jump,  reaches $10 ^{-6}$. The
VEPP-4M-KEDR collaboration has a lot of experience in using this method to
measure the masses of various particles~\cite{MRD_NPPP_2016}. In particular,
the masses of $J /\psi $ and $\psi (2S)$ were measured with an accuracy
better than the world average.

The other way is to observe the process of relaxation of the rate of counting
particles under conditions when a relatively slow depolarization of the beam
occurs because of internal factors, not because of a resonant external influence.
In our case, such a factor is quantum fluctuations in the presence of
noticeable stationary perturbations of the guide field. Below we describe this
method in more details.

The relaxation time of polarization (`polarization lifetime') is measured from
the time evolution of the  Touschek particle counting rate. Correct
determination of the polarization lifetime takes into account the Touschek beam
lifetime, the spin dependence of the IBS, and scattering on the residual gas by
solving the equation for the beam particle population $N(t)$:
\begin{equation}
	-	\frac{dN}{dt}  = \frac{1}{\tau_{\rm tsh}} \frac{N^2(t)}{N(0)} \frac{V(0)}{V(t)} \bigl(1-\delta(t)\bigr)+ \frac{N(t)}{\tau_{\rm bg}}. 
	\label{eq:eq4}
\end{equation}
Here, the  first term   corresponds to the IBS and $\tau_{\rm tsh}$ is the
characteristic Touschek beam lifetime; the second term  with $\tau_{\rm bg}$,
the characteristic background lifetime, describes background scattering on the
residual gas; $\delta(t)$ is the polarization contribution to IBS, proportional
to the square of the degree of polarization; $V(t)$ is the beam volume.

During the experiment, the  beam  volume $V$ (more precisely, the transverse
beam sizes because it is assumed that  the longitudinal size varies slightly),
the  beam currents $I_{\rm pol}$, $I_{\rm unpol}$  and the count rates $f_{\rm
pol}$, $f_{\rm unpol}$   for the first (polarized) and the second (unpolarized)
bunches are measured.  The relative count rate difference $\delta f(t) = f_{\rm pol}/f_{\rm
unpol}-1$ is calculated. These experimental data are fitted using the following
formulae, which are a solution to ($\ref{eq:eq4}$):
	\begin{equation}
		\begin{split}
      f_{i}(t) & =  \frac{p_{\rm tsh}}{\tau_{\rm tsh}} \frac{I_{i}^2(t)}{I_{i}(0) e f_{0}} 
     			\frac{1+\alpha_V I_{i}(0)}{ 1 + \alpha_V I_{i}(t)} 
			\bigl(1-\delta_{i}(t)\bigr)     
			 +   \frac{p_{\rm bg}}{\tau_{\rm bg}} \frac{I_{i}(t)}{e f_{\rm rev}} \\
			 I_{i}(t) & =      \frac{I_{i}(0) e^{-t/\tau_{\rm bg}}}{1 + (1 - e^{-t/\tau_{\rm bg}})\frac{\tau_{\rm bg}}{\tau_{\rm tsh}} -  \int_0^t e^{-t/\tau_{\rm bg}}(\delta_{i}(t)  + \delta V_i(t)) \frac{dt}{\tau_{\rm tsh}}} \\
       V(t)     &  =      V_0 \biggl( 1 + \frac{\alpha_V}{2}(I_{1} (t) + I_{2}(t)) \biggr) \hspace{0.1\textwidth}  \\
                         &\delta f(t) = \frac{f_{\rm pol}}{f_{\rm unpol}}-1		  \approx  -  \epsilon_1(t) e^{-t/\tau_{\rm bg}}\bigl[\delta_{\rm pol}(t) - \delta_{\rm unpol}(t)\bigr] +  \\
                & \quad + \frac{\epsilon(t,1)}{\epsilon(t,2)} \frac{\delta N + 
                \int_0^t e^{-t/\tau_{\rm bg}} [\delta_{\rm pol}(t)-\delta_{\rm unpol}(t)]dt/\tau_{\rm tsh}}{1+(1 - e^{-t/\tau_{\rm bg}})\tau_{\rm bg}/\tau_{\rm tsh}}.
		\end{split}
	\end{equation}
Here, the index $i$ marks the  polarized and  unpolarized bunches;
$\epsilon(t,j)$ is the factor taking into account  the relative probabilities
$p_{\rm tsh}$ and $p_{\rm bg}$ of registration of the Touschek particles and
the beam particles scattered by the residual gas, respectively:
\begin{equation*}
	\epsilon(t,j) =  \biggl [
\frac{p_{\rm bg}}{p_{\rm tsh}} \left(
1 + \frac{\tau_{\rm tsh}}{\tau_{\rm bg}} 
\right) +
\left(j - \frac{p_{\rm bg}}{p_{\rm tsh}}\right)
e^{-\frac{t}{\tau_{\rm bg}}} \biggr ]^{-1}; 
\end{equation*}
$ef_{0}$ is the current of a single electron; $\delta N =
I_{pol}(0)/I_{unpol}(0) - 1 $ is the relative difference in the particle
population of the bunches.  We introduce a time-dependent correction $\delta
V_i(t)$  to the volume of bunches, using its dependence on the current in the
linear approximation:
\begin{eqnarray}
	\delta V_i(t) & = & \alpha_V ( I_i(t) - I_i(0))  \approx \nonumber  
	\\ & \approx & - \alpha_V I_i(0)  \frac{(1 - e^{-t/\tau_{\rm bg}})(\tau_{\rm tsh} + \tau_{\rm bg})}{ \tau_{\rm tsh} + \tau_{\rm bg}(1 - e^{-t/\tau_{\rm bg}})}.
\end{eqnarray}

After the end of the acceleration process, the state of polarization of both
bunches changes with time.  In this connection, their polarization
contributions to the rate of counting the Touschek particles are described by
the equations
	\begin{equation}
		\begin{split}
      \delta_{\rm pol}(t) & =  \eta\left [ P e^{-t/\tau_d}  + P_{\infty} (1-e^{-t/\tau_d})\right ]^2  \\
      \delta_{\rm unpol}(t) & =  \eta\left [P_\infty (1-e^{-t/\tau_d})\right ]^2. 
		\end{split}
	\end{equation}
Here, $\eta$ is the Touschek polarization factor; $\tau_d$ is the radiative
depolarization time; $P$ is the residual degree of polarization of the
polarized bunch at the end  of acceleration, and $P_\infty = (8\sqrt{3}/15)\,
\tau_d/\tau_p\approx 4\cdot 10^{-3}$ is the equilibrium polarization degree at
$t\rightarrow \infty$.

The following free parameters are used for the fitting:
$\Delta = \eta  P^2 \approx 1\%$  is the polarization Touschek effect;  
$\tau_d $   is the polarization life time, which is an object of interest;
$\tau_{\rm tsh} \approx 8000$~s is the Touschek life time;
$\tau_{\rm bg} \approx 30\, 000$~s  is the background life time;
$p_{\rm tsh} \approx 0.2$    is the relative probability of registration of Touschek particles;
$p_{\rm bg} \approx 0.05$     is the relative probability of registration of particles scattered by the residual gas; 
$I_{\rm pol}(0)\approx I_{unpol}(0) \approx 2$~mA  are the initial currents of the first and second bunches, respectively;
$V_0$  is the initial  beam volume;
$\alpha_V\approx 0.1\% /{\rm mA}$ is the coefficient of dependence of the beam volume on the beam current. 

Time evolution of the measured quantity $\delta f(t)$ in the conditions when
the acceleration to  the target energy was just done,  but the compensating
solenoids stay switched off can be described as follows.  At the 'advance
energy',  the ratio of the bunch currents is adjusted to a level of $\delta
N\approx 1\div2\%$ because of the necessity to minimize the slope of the
dependence $\delta f (t)$ as a whole and the associated systematic error. For
this purpose,  we  kick out  portion by portion the redundant  bunch particles
using the VEPP-4M inflector. If  the bunch current ratio mentioned above  is
provided, then $\delta f(t)>0$ during all the time of observation.    The
depolarization process is enhanced during crossing of the critical energy area
and goes on after completion of the acceleration to the target energy.  By this
reason,  the quantity $\delta f(t)$ grows somewhat exponentially in the
positive direction. The characteristic time of this growth for a given target
energy is calculated (see Fig.~\ref{fig:Time}). The polarization in the beam
drops to zero and then another process becomes dominating ---  relaxation due
to the difference in the bunch currents.    Because of the difference in the
IBS beam lifetime,  the quantity $\delta f(t)$  begins to change in the
negative direction.  Asymptotically, it goes to zero.

\section{\label{sec:Results}Experimental results}
\begin{figure}[htb]
\centering
\includegraphics[width=85mm]{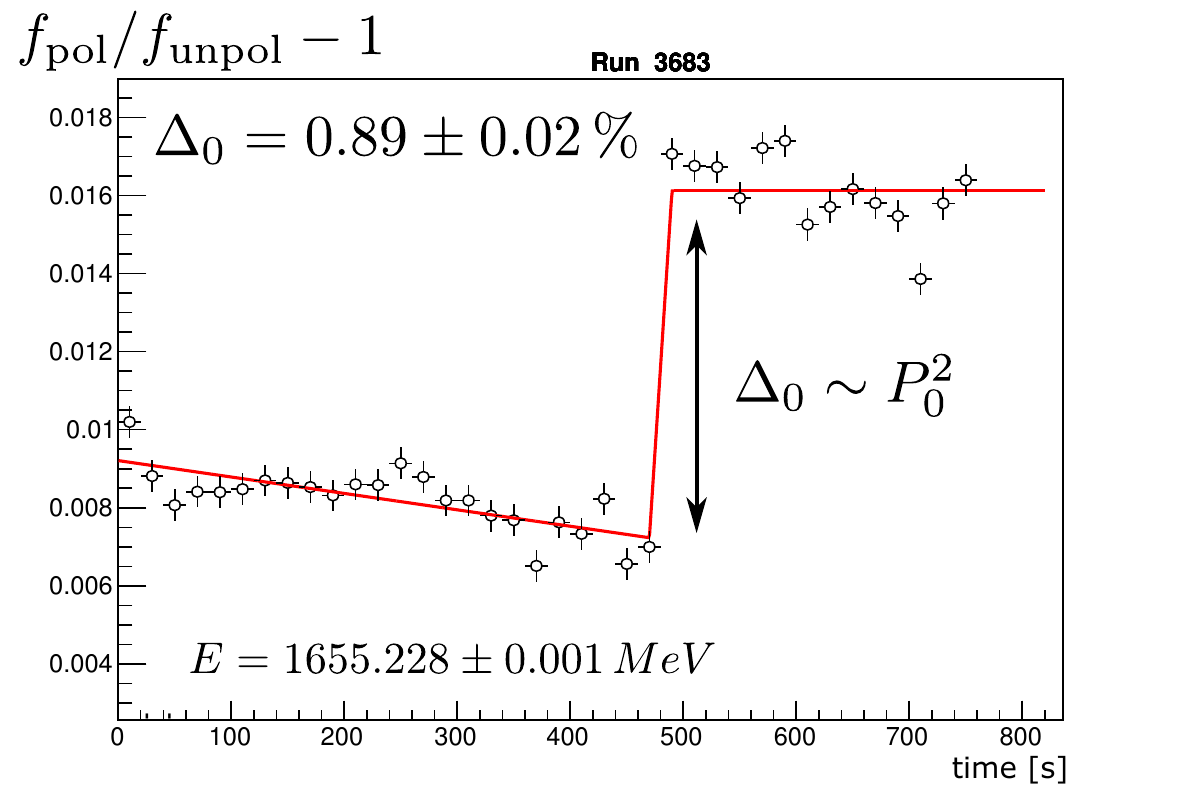}
(a)
\includegraphics[width=85mm]{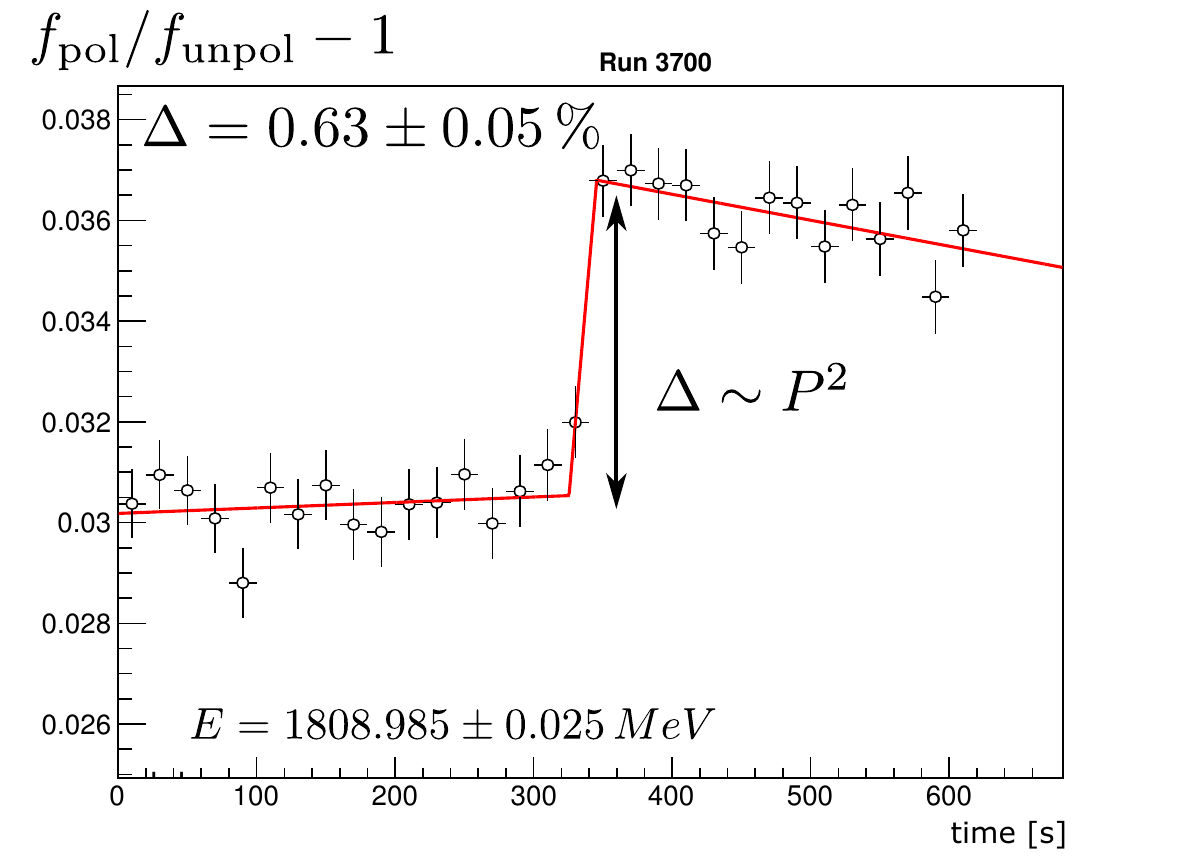}
(b)
\caption{Depolarization jumps  in two typical  scans  of the depolarizer
frequency: at the  advance energy 1655 MeV (a), i.e. before acceleration, and
at the target energy 1809 MeV after acceleration with a rate of 2.4~MeV/s (b).
In the second case, before scanning, the compensatory solenoid field was
restored in 385~s.  Each graph additionally shows the values of the measured
depolarization jump and the beam energy, as well as the corresponding errors.}
\label{fig:JmpAux}
\end{figure}

One of the RD beam energy calibrations  in the 'advance energy' mode is
presented in Fig.~\ref{fig:JmpAux}a. Basing on these data, one can get an idea
of the magnitude of the polarization effect measured with the Touschek
polarimeter. The time allotted for the radiative polarization at the VEPP-3
booster ring at the energy $E =1.65$~GeV was $5000\div6000$~s at the estimated
characteristic  time of polarization $\tau_p\approx4000$~s. On average, the
depolarization jump on the 'advance' energy was $\Delta_0=0.99 \pm 0.15~\%$.

\begin{figure}[htb]
\centering
\includegraphics[width=85mm]{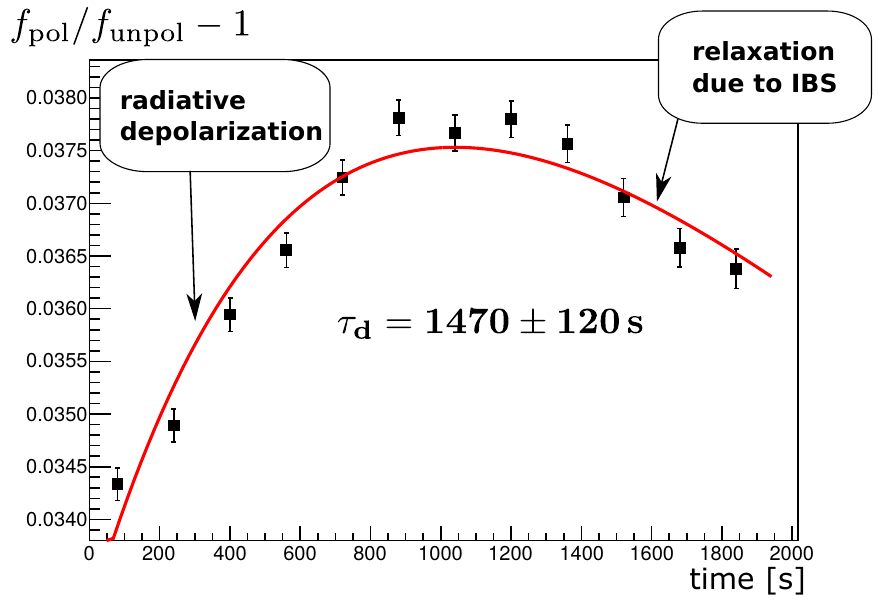}
\caption{The process of beam polarization relaxation following the acceleration
  from the 'advance' energy up to 1806~MeV with a rate of 5~MeV/s.  The
  compensatory solenoids remain switched off. In the time diagram of the ratio
  of the IBS rates in the polarized bunch  and unpolarized one, differing in
  the number of particles, two stages are clearly distinguished.}
\label{fig:Rlx}
\end{figure}

First of all, the effectiveness of the method was verified by observing the
process of beam  polarization relaxation (depolarization) after the
acceleration at the point E = 1.81~GeV was completed, but the anti-solenoids
remained switched off.  The observed relaxation may indicate that the
polarization in the beam is conserved by the end of the acceleration.  The time
evolution of the normalized Touschek electron counting rate is shown in
Fig.~\ref{fig:Rlx}.  In accordance with (\ref{eq:eq4}), the fit of the
experimental points takes into account the contributions of two processes. One
is  the radiative depolarization with the characteristic time $\tau_d$.  The
other is the  relaxation because of the Touschek losses of particles  provided
that the polarized and unpolarized bunches  are not equal in population.  In
the limit, the observed characteristic $f_{pol}/f_{unpol}-1$ tends to zero due
to the natural leveling of the bunches. Qualitatively, the relaxation process
proceeds as described in the previous section. The relaxation (depolarization)
time  $\tau_d=1470\pm120$ s determined from the data in Fig.~\ref{fig:Rlx}  is
in good agreement with the estimated time of about 1400 s (see
Fig.~\ref{fig:Time}).  

The fact of beam polarization preservation has been fully confirmed in the runs
on the beam energy measurement by the resonant depolarization technique applied
after the acceleration (Fig.~\ref{fig:JmpAux}b).

In contrast to the `spin relaxation' runs, the storage ring mode with the
anti-solenoid field switched on was restored before every start of the RD
procedure.  This measure stops the non-resonant (radiative) depolarization
process related to the contribution of strong longitudinal magnetic fields to
the spin-orbit coupling. An additional time of 385~s was required to restore
the anti-solenoid field   and, concurrently, to make necessary corrections to
the collider magnetic structure.  In accordance with the calculation, if the
acceleration stops at 1810 MeV with $P/P_0=0.84$, then the relative degree
falls down  to $0.77$ in 385~s.  In general,  the estimated degradation of the
depolarization jump $\Delta\sim P^2$ is  $(P/P_ 0)^2\approx0.6$ in magnitude. 

At the same time, the square root of the ratio of depolarization jumps measured
in several runs before and after acceleration is $\sqrt{\Delta/\Delta_0}\approx
0.76 \pm 0.12$. Taking into account the above additional degradation of
polarization during the anti-solenoid  field recovery time, the result of the
experiment to intersect an imperfection spin  resonance at a rate of 2.4~MeV/s
is characterized by the degree of polarization conservation
$P/P_0=0.83\pm0.13$.  Comparing all data in Fig.\ref{fig:Degree}, one can
conclude that the experiment and the calculation are in satisfactory
quantitative agreement.

\section{\label{sec:Discussion}Discussion}

The Partial Siberian Snake technique was first tested with protons at
IUCF~\cite{Krisch91} and is currently used in routine operation of BNL's
AGS~\cite{PhysRevLett.73.2982}. The specificity of its application to
electron-positron storage rings is the additional complication associated with
the depolarizing effect of synchrotron radiation.

Prior to our experiment at VEPP-4M, only one similar experiment was conducted,
at VEPP-2M~\cite{MRD_OMEGA_ENG}.  The radiative polarization was carried out at
600 MeV. Then the beam energy was lowered  to 380 MeV.  The polarization was
preserved due to the adiabatic crossing of the integer spin resonance at 440
MeV ($\nu=1$) at a rate of 10~MeV/s  using a solenoid described
in~\cite{Derb1977}. In our opinion, \cite{MRD_OMEGA_ENG} and our work differ
substantially in analysis of the experiment conditions.
In~\cite{MRD_OMEGA_ENG}, the authors considered the adiabatic condition as
decisive.  It was  briefly noted that the criterion for high-rate crossing in
relative to the rate of radiative depolarization was satisfied.  At the same
time, no attention was paid to the fact of the significant spin tune shift.
For the parameters of the solenoid at VEPP-2M, the angle of spin rotation in it
could be 0.2 radians. This yields a shift from the integer resonance of about
12 MeV.   Neither rigorous estimate of the radiative depolarization rate taking
into account changes of an equilibrium polarization direction, nor its
comparison with the experiment was made.  

An interesting experiment~\cite{NAKAMURA199893} on the acceleration of
polarized electrons from 1.2 to 3.5~GeV without using the partial Siberian
snake method was performed on the ELSA (Bonn Electron Stretcher Accelerator).
The polarization was measured in the extracted beams using the Moeller
polarimeter, which is sensitive to the sign of polarization. The ramping rate
was varied between 0.1 and 7~GeV/s, which is orders of magnitude greater than
rate possible with the non-laminated magnets at VEPP-4M.  The polarization was
preserved at acceleration  up to 1.9~GeV in the spin-flip (adiabatic) mode with
respect to the imperfection resonance  crossing.  According to the data
presented, in ELSA this should take place in the working range of the ramping
rate from 0.1 to 5~GeV/s.  In contrast to our case, due to the much higher
ramping rate, the difficulties in maintaining the polarization in the adiabatic
mode, caused by the influence of radiation and noted in section II, could be
avoided.
 
The complete Siberian Snake technique ($\varphi=\pi$) requires a much greater
integral of magnetic field: $4.6\nu$~Tesla$\times$meter.  The corresponding
spin tune shift is $1/2$,  or about 220~MeV. Thus, the detuning from the
integer spin resonances is maximum at any energy.  The radiative depolarization
factor is approximately $\tau_p/\tau_d\approx (11/54)\pi^2\nu^2$.  With
approaching to integer spin resonances, the Partial Siberian Snake solenoid
($\varphi \ll 1$) yields stronger spin-orbit coupling than the Siberian Snake
solenoid does. In accordance with (5) and (6),  the ratio of the respective
depolarization times is ($\nu=k$)
\begin{equation*}
\frac{\tau_d(\varphi=\pi)}{\tau_d (\varphi\ll 1)}\approx \frac{12}{\varphi ^2}.
\end{equation*}
In the VEPP-4M experiment, the longitudinal field integral is
$0.6\times3.3\approx 2$ Tesla$\times$meter against  $18.4$ Tesla$\times$meter
in  the Siberian Snake option.  The ratio of the depolarization times is about
$102$: $\tau_d(\varphi=\pi)=2.7$~h and  $\tau_d(\varphi=0.34$ rad)=97~s.  With
the distance from the integer resonance, the ratio of the relaxation times for
the cases of strong and weak solenoids is reversed in a qualitative sense.  To
see this, compare, for example, the curves in  Fig.\ref{fig:Time} for the
$50\,\% $ and $100\,\% $ decompensation of the detector field.  As applied to
our case, a complete Siberian snake could be better since it eliminates all
possible spin resonances and has a much smaller depolarizing effect at the
resonance energy.  At the same time, economically, the method based on
decompensation of the longitudinal field of the detector cost us nothing. With
its help we easily solved the practical problem that arose during our
experiment on high energy physics.

At VEPP-4M, the radiative depolarization can limit the effectiveness of the
method developed with growth of the target beam energy. Estimates show that it is possible
to intersect the third resonance $E=1322$~MeV practically without polarization
loss during beam deacceleration with a rate of 2 MeV/s starting from 1550~MeV.
However, acceleration in the same manner from 1.85 GeV up to 2.4~GeV, while crossing
the fifth resonance  $E=2203$~MeV,  will lead to a three-fold decrease in the
polarization degree. In this case, we plan to increase the ramping rate to
$10\div 20$~MeV per second. In addition,  it will be necessary to study the
possibilities of crossing the resonances $3-\nu_{x,z}$, $4+\nu_{x,z}$ and
$5-\nu_{x,z}$ falling within the specified ranges.

\bibliography{paper}

\providecommand{\noopsort}[1]{}\providecommand{\singleletter}[1]{#1}%
\begin{thebibliography}{32}%
\makeatletter
\providecommand \@ifxundefined [1]{%
 \@ifx{#1\undefined}
}%
\providecommand \@ifnum [1]{%
 \ifnum #1\expandafter \@firstoftwo
 \else \expandafter \@secondoftwo
 \fi
}%
\providecommand \@ifx [1]{%
 \ifx #1\expandafter \@firstoftwo
 \else \expandafter \@secondoftwo
 \fi
}%
\providecommand \natexlab [1]{#1}%
\providecommand \enquote  [1]{``#1''}%
\providecommand \bibnamefont  [1]{#1}%
\providecommand \bibfnamefont [1]{#1}%
\providecommand \citenamefont [1]{#1}%
\providecommand \href@noop [0]{\@secondoftwo}%
\providecommand \href [0]{\begingroup \@sanitize@url \@href}%
\providecommand \@href[1]{\@@startlink{#1}\@@href}%
\providecommand \@@href[1]{\endgroup#1\@@endlink}%
\providecommand \@sanitize@url [0]{\catcode `\\12\catcode `\$12\catcode
  `\&12\catcode `\#12\catcode `\^12\catcode `\_12\catcode `\%12\relax}%
\providecommand \@@startlink[1]{}%
\providecommand \@@endlink[0]{}%
\providecommand \url  [0]{\begingroup\@sanitize@url \@url }%
\providecommand \@url [1]{\endgroup\@href {#1}{\urlprefix }}%
\providecommand \urlprefix  [0]{URL }%
\providecommand \Eprint [0]{\href }%
\providecommand \doibase [0]{http://dx.doi.org/}%
\providecommand \selectlanguage [0]{\@gobble}%
\providecommand \bibinfo  [0]{\@secondoftwo}%
\providecommand \bibfield  [0]{\@secondoftwo}%
\providecommand \translation [1]{[#1]}%
\providecommand \BibitemOpen [0]{}%
\providecommand \bibitemStop [0]{}%
\providecommand \bibitemNoStop [0]{.\EOS\space}%
\providecommand \EOS [0]{\spacefactor3000\relax}%
\providecommand \BibitemShut  [1]{\csname bibitem#1\endcsname}%
\let\auto@bib@innerbib\@empty
\bibitem [{\citenamefont {Anashin}\ \emph {et~al.}(2013)\citenamefont
  {Anashin}, \citenamefont {Aulchenko}, \citenamefont {Baldin}, \citenamefont
  {Barladyan}, \citenamefont {Barnyakov}, \citenamefont {Barnyakov},
  \citenamefont {Baru}, \citenamefont {Basok}, \citenamefont {Bedny},
  \citenamefont {Beloborodova}, \citenamefont {Blinov}, \citenamefont {Blinov},
  \citenamefont {Bobrov}, \citenamefont {Bobrovnikov}, \citenamefont {Bondar},
  \citenamefont {Buzykaev}, \citenamefont {Vorobiov}, \citenamefont {Gulevich},
  \citenamefont {Dneprovsky}, \citenamefont {Zhilich}, \citenamefont
  {Zhulanov}, \citenamefont {Karpov}, \citenamefont {Karpov}, \citenamefont
  {Kononov}, \citenamefont {Kotov}, \citenamefont {Kravchenko}, \citenamefont
  {Kudryavtsev}, \citenamefont {Kuzmin}, \citenamefont {Kulikov}, \citenamefont
  {Kuper}, \citenamefont {Levichev}, \citenamefont {Maksimov}, \citenamefont
  {Malyshev}, \citenamefont {Maslennikov}, \citenamefont {Medvedko},
  \citenamefont {Muchnoi}, \citenamefont {Nikitin}, \citenamefont {Nikolaev},
  \citenamefont {Onuchin}, \citenamefont {Oreshkin}, \citenamefont {Orlov},
  \citenamefont {Osipov}, \citenamefont {Peleganchuk}, \citenamefont
  {Pivovarov}, \citenamefont {Poluektov}, \citenamefont {Pospelov},
  \citenamefont {Prisekin}, \citenamefont {Rodyakin}, \citenamefont {Ruban},
  \citenamefont {Savinov}, \citenamefont {Skovpen}, \citenamefont {Skrinsky},
  \citenamefont {Smalyuk}, \citenamefont {Snopkov}, \citenamefont {Sokolov},
  \citenamefont {Sukharev}, \citenamefont {Talyshev}, \citenamefont {Tayursky},
  \citenamefont {Telnov}, \citenamefont {Tikhonov}, \citenamefont {Todyshev},
  \citenamefont {Usov}, \citenamefont {Kharlamova}, \citenamefont {Shamov},
  \citenamefont {Shwartz}, \citenamefont {Shekhtman}, \citenamefont
  {Shusharo},\ and\ \citenamefont {Yushkov}}]{KEDR}%
  \BibitemOpen
  \bibfield  {author} {\bibinfo {author} {\bibfnamefont {V.}~\bibnamefont
  {Anashin}}, \bibinfo {author} {\bibfnamefont {V.}~\bibnamefont {Aulchenko}},
  \bibinfo {author} {\bibfnamefont {E.}~\bibnamefont {Baldin}}, \bibinfo
  {author} {\bibfnamefont {A.}~\bibnamefont {Barladyan}}, \bibinfo {author}
  {\bibfnamefont {A.}~\bibnamefont {Barnyakov}}, \bibinfo {author}
  {\bibfnamefont {M.}~\bibnamefont {Barnyakov}}, \bibinfo {author}
  {\bibfnamefont {S.}~\bibnamefont {Baru}}, \bibinfo {author} {\bibfnamefont
  {I.}~\bibnamefont {Basok}}, \bibinfo {author} {\bibfnamefont
  {I.}~\bibnamefont {Bedny}}, \bibinfo {author} {\bibfnamefont
  {O.}~\bibnamefont {Beloborodova}}, \bibinfo {author} {\bibfnamefont
  {A.}~\bibnamefont {Blinov}}, \bibinfo {author} {\bibfnamefont
  {V.}~\bibnamefont {Blinov}}, \bibinfo {author} {\bibfnamefont
  {A.}~\bibnamefont {Bobrov}}, \bibinfo {author} {\bibfnamefont
  {V.}~\bibnamefont {Bobrovnikov}}, \bibinfo {author} {\bibfnamefont
  {A.}~\bibnamefont {Bondar}}, \bibinfo {author} {\bibfnamefont
  {A.}~\bibnamefont {Buzykaev}}, \bibinfo {author} {\bibfnamefont
  {A.}~\bibnamefont {Vorobiov}}, \bibinfo {author} {\bibfnamefont
  {V.}~\bibnamefont {Gulevich}}, \bibinfo {author} {\bibfnamefont
  {L.}~\bibnamefont {Dneprovsky}}, \bibinfo {author} {\bibfnamefont
  {V.}~\bibnamefont {Zhilich}}, \bibinfo {author} {\bibfnamefont
  {V.}~\bibnamefont {Zhulanov}}, \bibinfo {author} {\bibfnamefont
  {G.}~\bibnamefont {Karpov}}, \bibinfo {author} {\bibfnamefont
  {S.}~\bibnamefont {Karpov}}, \bibinfo {author} {\bibfnamefont
  {S.}~\bibnamefont {Kononov}}, \bibinfo {author} {\bibfnamefont
  {K.}~\bibnamefont {Kotov}}, \bibinfo {author} {\bibfnamefont
  {E.}~\bibnamefont {Kravchenko}}, \bibinfo {author} {\bibfnamefont
  {V.}~\bibnamefont {Kudryavtsev}}, \bibinfo {author} {\bibfnamefont
  {A.}~\bibnamefont {Kuzmin}}, \bibinfo {author} {\bibfnamefont
  {V.}~\bibnamefont {Kulikov}}, \bibinfo {author} {\bibfnamefont
  {E.}~\bibnamefont {Kuper}}, \bibinfo {author} {\bibfnamefont
  {E.}~\bibnamefont {Levichev}}, \bibinfo {author} {\bibfnamefont
  {D.}~\bibnamefont {Maksimov}}, \bibinfo {author} {\bibfnamefont
  {V.}~\bibnamefont {Malyshev}}, \bibinfo {author} {\bibfnamefont
  {A.}~\bibnamefont {Maslennikov}}, \bibinfo {author} {\bibfnamefont
  {A.}~\bibnamefont {Medvedko}}, \bibinfo {author} {\bibfnamefont
  {N.}~\bibnamefont {Muchnoi}}, \bibinfo {author} {\bibfnamefont
  {S.}~\bibnamefont {Nikitin}}, \bibinfo {author} {\bibfnamefont
  {I.}~\bibnamefont {Nikolaev}}, \bibinfo {author} {\bibfnamefont
  {A.}~\bibnamefont {Onuchin}}, \bibinfo {author} {\bibfnamefont
  {S.}~\bibnamefont {Oreshkin}}, \bibinfo {author} {\bibfnamefont
  {I.}~\bibnamefont {Orlov}}, \bibinfo {author} {\bibfnamefont
  {A.}~\bibnamefont {Osipov}}, \bibinfo {author} {\bibfnamefont
  {S.}~\bibnamefont {Peleganchuk}}, \bibinfo {author} {\bibfnamefont
  {S.}~\bibnamefont {Pivovarov}}, \bibinfo {author} {\bibfnamefont
  {A.}~\bibnamefont {Poluektov}}, \bibinfo {author} {\bibfnamefont
  {G.}~\bibnamefont {Pospelov}}, \bibinfo {author} {\bibfnamefont
  {V.}~\bibnamefont {Prisekin}}, \bibinfo {author} {\bibfnamefont
  {V.}~\bibnamefont {Rodyakin}}, \bibinfo {author} {\bibfnamefont
  {A.}~\bibnamefont {Ruban}}, \bibinfo {author} {\bibfnamefont
  {G.}~\bibnamefont {Savinov}}, \bibinfo {author} {\bibfnamefont
  {Y.}~\bibnamefont {Skovpen}}, \bibinfo {author} {\bibfnamefont
  {A.}~\bibnamefont {Skrinsky}}, \bibinfo {author} {\bibfnamefont
  {V.}~\bibnamefont {Smalyuk}}, \bibinfo {author} {\bibfnamefont
  {R.}~\bibnamefont {Snopkov}}, \bibinfo {author} {\bibfnamefont
  {A.}~\bibnamefont {Sokolov}}, \bibinfo {author} {\bibfnamefont
  {A.}~\bibnamefont {Sukharev}}, \bibinfo {author} {\bibfnamefont
  {A.}~\bibnamefont {Talyshev}}, \bibinfo {author} {\bibfnamefont
  {V.}~\bibnamefont {Tayursky}}, \bibinfo {author} {\bibfnamefont
  {V.}~\bibnamefont {Telnov}}, \bibinfo {author} {\bibfnamefont
  {Y.}~\bibnamefont {Tikhonov}}, \bibinfo {author} {\bibfnamefont
  {K.}~\bibnamefont {Todyshev}}, \bibinfo {author} {\bibfnamefont
  {Y.}~\bibnamefont {Usov}}, \bibinfo {author} {\bibfnamefont {T.}~\bibnamefont
  {Kharlamova}}, \bibinfo {author} {\bibfnamefont {A.}~\bibnamefont {Shamov}},
  \bibinfo {author} {\bibfnamefont {B.}~\bibnamefont {Shwartz}}, \bibinfo
  {author} {\bibfnamefont {L.}~\bibnamefont {Shekhtman}}, \bibinfo {author}
  {\bibfnamefont {A.}~\bibnamefont {Shusharo}}, \ and\ \bibinfo {author}
  {\bibfnamefont {A.}~\bibnamefont {Yushkov}},\ }\href {\doibase
  10.1134/S1063779613040035} {\bibfield  {journal} {\bibinfo  {journal}
  {Physics of Particles and Nuclei}\ }\textbf {\bibinfo {volume} {44}},\
  \bibinfo {pages} {657} (\bibinfo {year} {2013})}\BibitemShut {NoStop}%
\bibitem [{\citenamefont {Anachin}\ \emph {et~al.}(1998)\citenamefont
  {Anachin}, \citenamefont {Anchugov}, \citenamefont {Bondar}, \citenamefont
  {Dubrovin}, \citenamefont {Durnov}, \citenamefont {Fomin}, \citenamefont
  {Y.}, \citenamefont {Gorniker}, \citenamefont {Kalinin}, \citenamefont
  {Karnaev}, \citenamefont {Kezerashvili}, \citenamefont {Kiselev},
  \citenamefont {Kuper}, \citenamefont {Kurkin}, \citenamefont {Y.},
  \citenamefont {Levichev}, \citenamefont {Medvedko}, \citenamefont
  {Mironenko}, \citenamefont {Mishnev}, \citenamefont {Muchnoi}, \citenamefont
  {Naumenkov}, \citenamefont {V.}, \citenamefont {Nikitin}, \citenamefont
  {Onuchin}, \citenamefont {Petrov}, \citenamefont {M.Petrov}, \citenamefont
  {V.Petrov}, \citenamefont {Popov}, \citenamefont {Protopopov}, \citenamefont
  {Y.}, \citenamefont {Shatilov}, \citenamefont {Simonov}, \citenamefont
  {Skrinsky}, \citenamefont {Smaluk}, \citenamefont {Tikhonov}, \citenamefont
  {Tumaikin}, \citenamefont {Zinevich},\ and\ \citenamefont {E.}}]{VEPP-4M}%
  \BibitemOpen
  \bibfield  {author} {\bibinfo {author} {\bibfnamefont {V.}~\bibnamefont
  {Anachin}}, \bibinfo {author} {\bibfnamefont {O.}~\bibnamefont {Anchugov}},
  \bibinfo {author} {\bibfnamefont {A.}~\bibnamefont {Bondar}}, \bibinfo
  {author} {\bibfnamefont {A.}~\bibnamefont {Dubrovin}}, \bibinfo {author}
  {\bibfnamefont {P.}~\bibnamefont {Durnov}}, \bibinfo {author} {\bibfnamefont
  {M.}~\bibnamefont {Fomin}}, \bibinfo {author} {\bibfnamefont
  {G.}~\bibnamefont {Y.}}, \bibinfo {author} {\bibfnamefont {E.}~\bibnamefont
  {Gorniker}}, \bibinfo {author} {\bibfnamefont {A.}~\bibnamefont {Kalinin}},
  \bibinfo {author} {\bibfnamefont {S.}~\bibnamefont {Karnaev}}, \bibinfo
  {author} {\bibfnamefont {G.}~\bibnamefont {Kezerashvili}}, \bibinfo {author}
  {\bibfnamefont {V.}~\bibnamefont {Kiselev}}, \bibinfo {author} {\bibfnamefont
  {E.}~\bibnamefont {Kuper}}, \bibinfo {author} {\bibfnamefont
  {G.}~\bibnamefont {Kurkin}}, \bibinfo {author} {\bibfnamefont
  {L.}~\bibnamefont {Y.}}, \bibinfo {author} {\bibfnamefont {B.}~\bibnamefont
  {Levichev}}, \bibinfo {author} {\bibfnamefont {A.}~\bibnamefont {Medvedko}},
  \bibinfo {author} {\bibfnamefont {L.}~\bibnamefont {Mironenko}}, \bibinfo
  {author} {\bibfnamefont {S.}~\bibnamefont {Mishnev}}, \bibinfo {author}
  {\bibfnamefont {N.}~\bibnamefont {Muchnoi}}, \bibinfo {author} {\bibfnamefont
  {A.}~\bibnamefont {Naumenkov}}, \bibinfo {author} {\bibfnamefont
  {N.}~\bibnamefont {V.}}, \bibinfo {author} {\bibfnamefont {S.}~\bibnamefont
  {Nikitin}}, \bibinfo {author} {\bibfnamefont {A.}~\bibnamefont {Onuchin}},
  \bibinfo {author} {\bibfnamefont {S.}~\bibnamefont {Petrov}}, \bibinfo
  {author} {\bibfnamefont {V.}~\bibnamefont {M.Petrov}}, \bibinfo {author}
  {\bibfnamefont {V.}~\bibnamefont {V.Petrov}}, \bibinfo {author}
  {\bibfnamefont {V.}~\bibnamefont {Popov}}, \bibinfo {author} {\bibfnamefont
  {I.}~\bibnamefont {Protopopov}}, \bibinfo {author} {\bibfnamefont
  {P.}~\bibnamefont {Y.}}, \bibinfo {author} {\bibfnamefont {D.}~\bibnamefont
  {Shatilov}}, \bibinfo {author} {\bibfnamefont {E.}~\bibnamefont {Simonov}},
  \bibinfo {author} {\bibfnamefont {A.}~\bibnamefont {Skrinsky}}, \bibinfo
  {author} {\bibfnamefont {V.}~\bibnamefont {Smaluk}}, \bibinfo {author}
  {\bibfnamefont {Y.}~\bibnamefont {Tikhonov}}, \bibinfo {author}
  {\bibfnamefont {G.}~\bibnamefont {Tumaikin}}, \bibinfo {author}
  {\bibfnamefont {N.}~\bibnamefont {Zinevich}}, \ and\ \bibinfo {author}
  {\bibfnamefont {Z.}~\bibnamefont {E.}},\ }in\ \href@noop {} {\emph {\bibinfo
  {booktitle} {{6th European Particle Accelerator Conference (EPAC 98)}}}}\
  (\bibinfo {address} {{Stockholm, Sweden}},\ \bibinfo {year} {1998})\ pp.\
  \bibinfo {pages} {400--402}\BibitemShut {NoStop}%
\bibitem [{\citenamefont {Dyug}\ \emph {et~al.}(2005)\citenamefont {Dyug},
  \citenamefont {Grigoriev}, \citenamefont {Kiselev}, \citenamefont
  {Lazarenko}, \citenamefont {Levichev}, \citenamefont {Mikaiylov},
  \citenamefont {Mishnev}, \citenamefont {Nikitin}, \citenamefont {Nikolenko},
  \citenamefont {Rachek}, \citenamefont {Shestakov}, \citenamefont {Toporkov},
  \citenamefont {Zevakov},\ and\ \citenamefont {Zhilich}}]{DYUG}%
  \BibitemOpen
  \bibfield  {author} {\bibinfo {author} {\bibfnamefont {M.}~\bibnamefont
  {Dyug}}, \bibinfo {author} {\bibfnamefont {A.}~\bibnamefont {Grigoriev}},
  \bibinfo {author} {\bibfnamefont {V.}~\bibnamefont {Kiselev}}, \bibinfo
  {author} {\bibfnamefont {B.}~\bibnamefont {Lazarenko}}, \bibinfo {author}
  {\bibfnamefont {E.}~\bibnamefont {Levichev}}, \bibinfo {author}
  {\bibfnamefont {A.}~\bibnamefont {Mikaiylov}}, \bibinfo {author}
  {\bibfnamefont {S.}~\bibnamefont {Mishnev}}, \bibinfo {author} {\bibfnamefont
  {S.}~\bibnamefont {Nikitin}}, \bibinfo {author} {\bibfnamefont
  {D.}~\bibnamefont {Nikolenko}}, \bibinfo {author} {\bibfnamefont
  {I.}~\bibnamefont {Rachek}}, \bibinfo {author} {\bibfnamefont
  {Y.}~\bibnamefont {Shestakov}}, \bibinfo {author} {\bibfnamefont
  {D.}~\bibnamefont {Toporkov}}, \bibinfo {author} {\bibfnamefont
  {S.}~\bibnamefont {Zevakov}}, \ and\ \bibinfo {author} {\bibfnamefont
  {V.}~\bibnamefont {Zhilich}},\ }\href {\doibase
  http://dx.doi.org/10.1016/j.nima.2004.08.095} {\bibfield  {journal} {\bibinfo
   {journal} {Nuclear Instruments and Methods in Physics Research Section A:
  Accelerators, Spectrometers, Detectors and Associated Equipment}\ }\textbf
  {\bibinfo {volume} {536}},\ \bibinfo {pages} {338 } (\bibinfo {year}
  {2005})},\ \bibinfo {note} {polarized Sources and Targets for the 21st
  Century. Proceedings o f the 10th International Workshop on Polarized Sources
  and Targets}\BibitemShut {NoStop}%
\bibitem [{\citenamefont {Anashin}\ \emph {et~al.}(2007)\citenamefont
  {Anashin}, \citenamefont {Aulchenko}, \citenamefont {Nikolaev} \emph
  {et~al.}}]{TAU}%
  \BibitemOpen
  \bibfield  {author} {\bibinfo {author} {\bibfnamefont {V.~V.}\ \bibnamefont
  {Anashin}}, \bibinfo {author} {\bibfnamefont {V.}~\bibnamefont {Aulchenko}},
  \bibinfo {author} {\bibfnamefont {I.}~\bibnamefont {Nikolaev}},  \emph
  {et~al.},\ }\href {\doibase 10.1134/S0021364007080012} {\bibfield  {journal}
  {\bibinfo  {journal} {{JETP Lett.}}\ }\textbf {\bibinfo {volume} {85}},\
  \bibinfo {pages} {347} (\bibinfo {year} {2007})}\BibitemShut {NoStop}%
\bibitem [{\citenamefont {Froissart}\ and\ \citenamefont
  {Stora}(1960)}]{Froissart:1960zz}%
  \BibitemOpen
  \bibfield  {author} {\bibinfo {author} {\bibfnamefont {M.}~\bibnamefont
  {Froissart}}\ and\ \bibinfo {author} {\bibfnamefont {R.}~\bibnamefont
  {Stora}},\ }\href {\doibase 10.1016/0029-554X(60)90033-1} {\bibfield
  {journal} {\bibinfo  {journal} {Nucl. Instrum. Meth.}\ }\textbf {\bibinfo
  {volume} {7}},\ \bibinfo {pages} {297} (\bibinfo {year} {1960})}\BibitemShut
  {NoStop}%
\bibitem [{\citenamefont {Derbenev}\ and\ \citenamefont
  {Kondratenko}(1974)}]{DERB}%
  \BibitemOpen
  \bibfield  {author} {\bibinfo {author} {\bibfnamefont {Y.}~\bibnamefont
  {Derbenev}}\ and\ \bibinfo {author} {\bibfnamefont {A.}~\bibnamefont
  {Kondratenko}},\ }\href@noop {} {\bibfield  {journal} {\bibinfo  {journal}
  {{DAN SSSR}}\ }\textbf {\bibinfo {volume} {217}},\ \bibinfo {pages} {311}
  (\bibinfo {year} {1974})}\BibitemShut {NoStop}%
\bibitem [{\citenamefont {Bogomyagkov}\ \emph {et~al.}(2004)\citenamefont
  {Bogomyagkov}, \citenamefont {Nikitin}, \citenamefont {Nikolaev},
  \citenamefont {Kiselev}, \citenamefont {Kremyanskaya}, \citenamefont
  {Levichev}, \citenamefont {Simonov},\ and\ \citenamefont
  {Skrinsky}}]{TAU-POL-EPAC-2004}%
  \BibitemOpen
  \bibfield  {author} {\bibinfo {author} {\bibfnamefont {A.}~\bibnamefont
  {Bogomyagkov}}, \bibinfo {author} {\bibfnamefont {S.}~\bibnamefont
  {Nikitin}}, \bibinfo {author} {\bibfnamefont {I.}~\bibnamefont {Nikolaev}},
  \bibinfo {author} {\bibfnamefont {V.}~\bibnamefont {Kiselev}}, \bibinfo
  {author} {\bibfnamefont {E.}~\bibnamefont {Kremyanskaya}}, \bibinfo {author}
  {\bibfnamefont {E.}~\bibnamefont {Levichev}}, \bibinfo {author}
  {\bibfnamefont {E.}~\bibnamefont {Simonov}}, \ and\ \bibinfo {author}
  {\bibfnamefont {A.}~\bibnamefont {Skrinsky}},\ }in\ \href@noop {} {\emph
  {\bibinfo {booktitle} {{9th European Particle Accelerator Conference (EPAC
  2004)}}}}\ (\bibinfo {address} {{Lucerne, Switzerland}},\ \bibinfo {year}
  {2004})\BibitemShut {NoStop}%
\bibitem [{\citenamefont {Derbenev}\ \emph {et~al.}(1979)\citenamefont
  {Derbenev}, \citenamefont {Kondratenko},\ and\ \citenamefont
  {Skrinsky}}]{DERB2}%
  \BibitemOpen
  \bibfield  {author} {\bibinfo {author} {\bibfnamefont {Y.~S.}\ \bibnamefont
  {Derbenev}}, \bibinfo {author} {\bibfnamefont {A.~M.}\ \bibnamefont
  {Kondratenko}}, \ and\ \bibinfo {author} {\bibfnamefont {A.~N.}\ \bibnamefont
  {Skrinsky}},\ }\href@noop {} {\bibfield  {journal} {\bibinfo  {journal}
  {{Part. Accel.}}\ }\textbf {\bibinfo {volume} {9}},\ \bibinfo {pages} {247}
  (\bibinfo {year} {1979})}\BibitemShut {NoStop}%
\bibitem [{\citenamefont {Derbenev}\ and\ \citenamefont
  {Kondratenko}(1973)}]{Derbenev:1973ia}%
  \BibitemOpen
  \bibfield  {author} {\bibinfo {author} {\bibfnamefont {Y.}~\bibnamefont
  {Derbenev}}\ and\ \bibinfo {author} {\bibfnamefont {A.}~\bibnamefont
  {Kondratenko}},\ }\href@noop {} {\bibfield  {journal} {\bibinfo  {journal}
  {Sov.Phys.JETP}\ }\textbf {\bibinfo {volume} {37}},\ \bibinfo {pages} {968}
  (\bibinfo {year} {1973})}\BibitemShut {NoStop}%
\bibitem [{\citenamefont {Sokolov}\ and\ \citenamefont
  {Ternov}(1964)}]{SokolovTernov}%
  \BibitemOpen
  \bibfield  {author} {\bibinfo {author} {\bibfnamefont {A.}~\bibnamefont
  {Sokolov}}\ and\ \bibinfo {author} {\bibfnamefont {I.}~\bibnamefont
  {Ternov}},\ }\href@noop {} {\bibfield  {journal} {\bibinfo  {journal} {{Sov.
  Phys. Dokl}}\ }\textbf {\bibinfo {volume} {8}},\ \bibinfo {pages} {1203}
  (\bibinfo {year} {1964})}\BibitemShut {NoStop}%
\bibitem [{\citenamefont {{Kondratenko, A.M.}}(1982)}]{KondratenkoDT}%
  \BibitemOpen
  \bibfield  {author} {\bibinfo {author} {\bibnamefont {{Kondratenko, A.M.}}},\
  }\href@noop {} {\enquote {\bibinfo {title} {{Polarized beams in storage rings
  and cyclic accelerators}},}\ } (\bibinfo {year} {{1982}}),\ \bibinfo {note}
  {{Doctoral thesis, Budker Institute of Nuclear Physics, Novosibirsk, USSR (in
  russian)}}\BibitemShut {NoStop}%
\bibitem [{\citenamefont {Nikitin}(2015)}]{Nikitin-BINP-2015-1}%
  \BibitemOpen
  \bibfield  {author} {\bibinfo {author} {\bibfnamefont {S.}~\bibnamefont
  {Nikitin}},\ }\href@noop {} {\bibfield  {journal} {\bibinfo  {journal}
  {{Budker Institute of Nuclear Physics, Novosibirsk, preprint}}\ }\textbf
  {\bibinfo {volume} {2015-1}} (\bibinfo {year} {2015})}\BibitemShut {NoStop}%
\bibitem [{\citenamefont {Barladyan}\ \emph {et~al.}(2017)\citenamefont
  {Barladyan} \emph {et~al.}}]{Barladyan:IPAC-17}%
  \BibitemOpen
  \bibfield  {author} {\bibinfo {author} {\bibfnamefont {A.}~\bibnamefont
  {Barladyan}} \emph {et~al.},\ }in\ \href {\doibase
  https://doi.org/10.18429/JACoW-IPAC2017-THPVA017} {\emph {\bibinfo
  {booktitle} {Proc. of International Particle Accelerator Conference
  (IPAC'17), Copenhagen, Denmark, 14---19 May, 2017}}},\ \bibinfo {series and
  number} {\bibinfo {series} {International Particle Accelerator Conference}\
  No.~\bibinfo {number} {8}}\ (\bibinfo  {publisher} {JACoW},\ \bibinfo
  {address} {Geneva, Switzerland},\ \bibinfo {year} {2017})\ pp.\ \bibinfo
  {pages} {4451--4453},\ \bibinfo {note}
  {https://doi.org/10.18429/JACoW-IPAC2017-THPVA017}\BibitemShut {NoStop}%
\bibitem [{\citenamefont {Derbenev}\ \emph {et~al.}(1977)\citenamefont
  {Derbenev}, \citenamefont {Kondratenko}, \citenamefont {Skrinsky},\ and\
  \citenamefont {Shatunov}}]{Derb1977}%
  \BibitemOpen
  \bibfield  {author} {\bibinfo {author} {\bibfnamefont {Y.}~\bibnamefont
  {Derbenev}}, \bibinfo {author} {\bibfnamefont {A.}~\bibnamefont
  {Kondratenko}}, \bibinfo {author} {\bibfnamefont {A.}~\bibnamefont
  {Skrinsky}}, \ and\ \bibinfo {author} {\bibfnamefont {Y.}~\bibnamefont
  {Shatunov}},\ }in\ \href@noop {} {\emph {\bibinfo {booktitle} {{proc. 10th
  Int'l Conference on High Energy Accelerators, Serpukhov}}}}\ (\bibinfo {year}
  {1977})\ pp.\ \bibinfo {pages} {76--80}\BibitemShut {NoStop}%
\bibitem [{\citenamefont {Derbenev}\ \emph {et~al.}(1978)\citenamefont
  {Derbenev}, \citenamefont {Kondratenko}, \citenamefont {Serednyakov},
  \citenamefont {Skrinsky}, \citenamefont {Tumaikin} \emph
  {et~al.}}]{Derbenev:1978hv}%
  \BibitemOpen
  \bibfield  {author} {\bibinfo {author} {\bibfnamefont {Y.}~\bibnamefont
  {Derbenev}}, \bibinfo {author} {\bibfnamefont {A.}~\bibnamefont
  {Kondratenko}}, \bibinfo {author} {\bibfnamefont {S.}~\bibnamefont
  {Serednyakov}}, \bibinfo {author} {\bibfnamefont {A.}~\bibnamefont
  {Skrinsky}}, \bibinfo {author} {\bibfnamefont {G.}~\bibnamefont {Tumaikin}},
  \emph {et~al.},\ }\href@noop {} {\bibfield  {journal} {\bibinfo  {journal}
  {Part.Accel.}\ }\textbf {\bibinfo {volume} {8}},\ \bibinfo {pages} {115}
  (\bibinfo {year} {1978})}\BibitemShut {NoStop}%
\bibitem [{\citenamefont {Nikitin}\ and\ \citenamefont
  {Saldin}(1981)}]{NikSaldin}%
  \BibitemOpen
  \bibfield  {author} {\bibinfo {author} {\bibfnamefont {S.}~\bibnamefont
  {Nikitin}}\ and\ \bibinfo {author} {\bibfnamefont {E.}~\bibnamefont
  {Saldin}},\ }\href@noop {} {\bibfield  {journal} {\bibinfo  {journal}
  {{preprint IYaF SO AN SSSR}}\ }\textbf {\bibinfo {volume} {1981-19}}
  (\bibinfo {year} {1981})},\ \bibinfo {note} {{Internal Report DESY L-Trans-
  290, Hamburg 1984}}\BibitemShut {NoStop}%
\bibitem [{\citenamefont {Blinov}\ \emph {et~al.}(2016)\citenamefont {Blinov}
  \emph {et~al.}}]{MRD_NPPP_2016}%
  \BibitemOpen
  \bibfield  {author} {\bibinfo {author} {\bibfnamefont {V.~E.}\ \bibnamefont
  {Blinov}} \emph {et~al.},\ }\bibfield  {booktitle} {\emph {\bibinfo
  {booktitle} {{Proceedings, 37th International Conference on High Energy
  Physics (ICHEP 2014): Valencia, Spain, July 2-9, 2014}}},\ }\href {\doibase
  10.1016/j.nuclphysbps.2015.09.028} {\bibfield  {journal} {\bibinfo  {journal}
  {{Nucl. Part. Phys. Proc.}}\ }\textbf {\bibinfo {volume} {273-275}},\
  \bibinfo {pages} {210} (\bibinfo {year} {2016})}\BibitemShut {NoStop}%
\bibitem [{\citenamefont {Nikitin}(1996)}]{NikCtau}%
  \BibitemOpen
  \bibfield  {author} {\bibinfo {author} {\bibfnamefont {S.}~\bibnamefont
  {Nikitin}},\ }\href {\doibase 10.1016/0168-9002(96)00529-3} {\bibfield
  {journal} {\bibinfo  {journal} {Nuclear Instruments and Methods in Physics
  Research A}\ }\textbf {\bibinfo {volume} {378}},\ \bibinfo {pages} {495}
  (\bibinfo {year} {1996})}\BibitemShut {NoStop}%
\bibitem [{\citenamefont {Nikitin}(2006)}]{NikRupac}%
  \BibitemOpen
  \bibfield  {author} {\bibinfo {author} {\bibfnamefont {S.}~\bibnamefont
  {Nikitin}},\ }in\ \href@noop {} {\emph {\bibinfo {booktitle} {{20th Russian
  Conference on Charged Particle Accelerators (RuPAC 2006)}}}}\ (\bibinfo
  {address} {{Novosibirsk, Russian Federation}},\ \bibinfo {year} {2006})\ p.\
  \bibinfo {pages} {MOAP01}\BibitemShut {NoStop}%
\bibitem [{\citenamefont {Steffen}(1982)}]{Steffen1982}%
  \BibitemOpen
  \bibfield  {author} {\bibinfo {author} {\bibfnamefont {K.}~\bibnamefont
  {Steffen}},\ }\href@noop {} {}\bibinfo {type} {Internal DESY report
  HERA82/11}\ (\bibinfo  {institution} {DESY},\ \bibinfo {year}
  {1982})\BibitemShut {NoStop}%
\bibitem [{\citenamefont {Nikitin}\ and\ \citenamefont
  {Protopopov}(1999)}]{NikProt1999}%
  \BibitemOpen
  \bibfield  {author} {\bibinfo {author} {\bibfnamefont {S.}~\bibnamefont
  {Nikitin}}\ and\ \bibinfo {author} {\bibfnamefont {I.}~\bibnamefont
  {Protopopov}},\ }\href@noop {} {\bibfield  {journal} {\bibinfo  {journal}
  {{preprint BINP SB RAS}}\ }\textbf {\bibinfo {volume} {1999-44}} (\bibinfo
  {year} {1999})}\BibitemShut {NoStop}%
\bibitem [{\citenamefont {Nikitin}\ and\ \citenamefont
  {Simonov}(2001)}]{Nikitin:2001ej}%
  \BibitemOpen
  \bibfield  {author} {\bibinfo {author} {\bibfnamefont {S.}~\bibnamefont
  {Nikitin}}\ and\ \bibinfo {author} {\bibfnamefont {E.}~\bibnamefont
  {Simonov}},\ }in\ \href@noop {} {\emph {\bibinfo {booktitle} {2nd Asian
  Particle Accelerator Conference (APAC 2001)}}}\ (\bibinfo {address} {Beijing,
  China},\ \bibinfo {year} {2001})\ pp.\ \bibinfo {pages}
  {457--459}\BibitemShut {NoStop}%
\bibitem [{\citenamefont {Blinov}\ \emph {et~al.}(2002)\citenamefont {Blinov},
  \citenamefont {Bogomyagkov}, \citenamefont {Nikolaev}, \citenamefont
  {Karnaev}, \citenamefont {Kiselev}, \citenamefont {Levichev}, \citenamefont
  {Levichev}, \citenamefont {Meshkov}, \citenamefont {Mishnev}, \citenamefont
  {Naumenkov}, \citenamefont {Nikitin}, \citenamefont {Popov}, \citenamefont
  {Polunin}, \citenamefont {Protopopov}, \citenamefont {Shatilov},
  \citenamefont {Simonov}, \citenamefont {Tikhonov},\ and\ \citenamefont
  {Tumaikin}}]{TSH-POL-NIM-2002}%
  \BibitemOpen
  \bibfield  {author} {\bibinfo {author} {\bibfnamefont {V.}~\bibnamefont
  {Blinov}}, \bibinfo {author} {\bibfnamefont {A.}~\bibnamefont {Bogomyagkov}},
  \bibinfo {author} {\bibfnamefont {I.}~\bibnamefont {Nikolaev}}, \bibinfo
  {author} {\bibfnamefont {S.}~\bibnamefont {Karnaev}}, \bibinfo {author}
  {\bibfnamefont {V.}~\bibnamefont {Kiselev}}, \bibinfo {author} {\bibfnamefont
  {B.}~\bibnamefont {Levichev}}, \bibinfo {author} {\bibfnamefont
  {E.}~\bibnamefont {Levichev}}, \bibinfo {author} {\bibfnamefont
  {O.}~\bibnamefont {Meshkov}}, \bibinfo {author} {\bibfnamefont
  {S.}~\bibnamefont {Mishnev}}, \bibinfo {author} {\bibfnamefont
  {A.}~\bibnamefont {Naumenkov}}, \bibinfo {author} {\bibfnamefont
  {S.}~\bibnamefont {Nikitin}}, \bibinfo {author} {\bibfnamefont
  {V.}~\bibnamefont {Popov}}, \bibinfo {author} {\bibfnamefont
  {A.}~\bibnamefont {Polunin}}, \bibinfo {author} {\bibfnamefont
  {I.}~\bibnamefont {Protopopov}}, \bibinfo {author} {\bibfnamefont
  {D.}~\bibnamefont {Shatilov}}, \bibinfo {author} {\bibfnamefont
  {E.}~\bibnamefont {Simonov}}, \bibinfo {author} {\bibfnamefont
  {Y.}~\bibnamefont {Tikhonov}}, \ and\ \bibinfo {author} {\bibfnamefont
  {G.}~\bibnamefont {Tumaikin}},\ }\href {\doibase
  10.1016/S0168-9002(02)01449-3} {\bibfield  {journal} {\bibinfo  {journal}
  {{Nucl. Instrum. Meth.}}\ }\textbf {\bibinfo {volume} {A494}},\ \bibinfo
  {pages} {81} (\bibinfo {year} {2002})}\BibitemShut {NoStop}%
\bibitem [{\citenamefont {Blinov}\ \emph {et~al.}(2009)\citenamefont {Blinov},
  \citenamefont {Bogomyagkov}, \citenamefont {Nikolaev}, \citenamefont
  {Muchnoi}, \citenamefont {Nikitin}, \citenamefont {Shamov},\ and\
  \citenamefont {Zhilich}}]{EMS-REVIEW-NIM-2009}%
  \BibitemOpen
  \bibfield  {author} {\bibinfo {author} {\bibfnamefont {V.~E.}\ \bibnamefont
  {Blinov}}, \bibinfo {author} {\bibfnamefont {A.~V.}\ \bibnamefont
  {Bogomyagkov}}, \bibinfo {author} {\bibfnamefont {I.~B.}\ \bibnamefont
  {Nikolaev}}, \bibinfo {author} {\bibfnamefont {N.~Y.}\ \bibnamefont
  {Muchnoi}}, \bibinfo {author} {\bibfnamefont {S.~A.}\ \bibnamefont
  {Nikitin}}, \bibinfo {author} {\bibfnamefont {A.~G.}\ \bibnamefont {Shamov}},
  \ and\ \bibinfo {author} {\bibfnamefont {V.~N.}\ \bibnamefont {Zhilich}},\
  }\href {\doibase 10.1016/j.nima.2008.08.078} {\bibfield  {journal} {\bibinfo
  {journal} {{Nucl. Instrum. Meth.}}\ }\textbf {\bibinfo {volume} {A598}},\
  \bibinfo {pages} {23} (\bibinfo {year} {2009})}\BibitemShut {NoStop}%
\bibitem [{\citenamefont {Bayer}\ and\ \citenamefont
  {Khoze}(1969)}]{bayer1969}%
  \BibitemOpen
  \bibfield  {author} {\bibinfo {author} {\bibfnamefont {V.}~\bibnamefont
  {Bayer}}\ and\ \bibinfo {author} {\bibfnamefont {V.}~\bibnamefont {Khoze}},\
  }\href@noop {} {\bibfield  {journal} {\bibinfo  {journal} {{Yadernaya
  Fizika}}\ }\textbf {\bibinfo {volume} {9}},\ \bibinfo {pages} {409} (\bibinfo
  {year} {1969})}\BibitemShut {NoStop}%
\bibitem [{\citenamefont {Bayer}\ \emph {et~al.}(1978)\citenamefont {Bayer},
  \citenamefont {Khoze},\ and\ \citenamefont {Strakhovenko}}]{BKS}%
  \BibitemOpen
  \bibfield  {author} {\bibinfo {author} {\bibfnamefont {V.}~\bibnamefont
  {Bayer}}, \bibinfo {author} {\bibfnamefont {V.}~\bibnamefont {Khoze}}, \ and\
  \bibinfo {author} {\bibfnamefont {V.}~\bibnamefont {Strakhovenko}},\
  }\href@noop {} {\bibfield  {journal} {\bibinfo  {journal} {{Doklada AN
  SSSR}}\ }\textbf {\bibinfo {volume} {241}},\ \bibinfo {pages} {797} (\bibinfo
  {year} {1978})}\BibitemShut {NoStop}%
\bibitem [{\citenamefont {Strakhovenko}(2011)}]{STRAKH}%
  \BibitemOpen
  \bibfield  {author} {\bibinfo {author} {\bibfnamefont {V.~M.}\ \bibnamefont
  {Strakhovenko}},\ }\href {\doibase 10.1103/PhysRevSTAB.14.012803} {\bibfield
  {journal} {\bibinfo  {journal} {{Phys. Rev. ST Accel. Beams}}\ }\textbf
  {\bibinfo {volume} {14}},\ \bibinfo {pages} {012803} (\bibinfo {year}
  {2011})},\ \Eprint {http://arxiv.org/abs/{0912.5429}} {{arXiv}:{0912.5429}
  [{physics.acc-ph}]} \BibitemShut {NoStop}%
\bibitem [{\citenamefont {Nikitin}\ and\ \citenamefont
  {Nikolaev}(2012)}]{TSH-CALC-JETP-2012}%
  \BibitemOpen
  \bibfield  {author} {\bibinfo {author} {\bibfnamefont {S.~A.}\ \bibnamefont
  {Nikitin}}\ and\ \bibinfo {author} {\bibfnamefont {I.~B.}\ \bibnamefont
  {Nikolaev}},\ }\href {\doibase {10.1134/S1063776112060106}} {\bibfield
  {journal} {\bibinfo  {journal} {{Journal of Experimental and Theoretical
  Physics}}\ }\textbf {\bibinfo {volume} {115}},\ \bibinfo {pages} {36}
  (\bibinfo {year} {2012})}\BibitemShut {NoStop}%
\bibitem [{\citenamefont {Roser}(1991)}]{Krisch91}%
  \BibitemOpen
  \bibfield  {author} {\bibinfo {author} {\bibfnamefont {T.}~\bibnamefont
  {Roser}},\ }in\ \href@noop {} {\emph {\bibinfo {booktitle} {High Energy Spin
  Physics}}},\ \bibinfo {editor} {edited by\ \bibinfo {editor} {\bibfnamefont
  {K.-H.}\ \bibnamefont {Althoff}}\ and\ \bibinfo {editor} {\bibfnamefont
  {W.}~\bibnamefont {Meyer}}}\ (\bibinfo  {publisher} {Springer Berlin
  Heidelberg},\ \bibinfo {address} {Berlin, Heidelberg},\ \bibinfo {year}
  {1991})\ pp.\ \bibinfo {pages} {284--291}\BibitemShut {NoStop}%
\bibitem [{\citenamefont {Huang}\ \emph {et~al.}(1994)\citenamefont {Huang},
  \citenamefont {Ahrens}, \citenamefont {Alessi}, \citenamefont {Beddo},
  \citenamefont {Brown}, \citenamefont {Bunce}, \citenamefont {Caussyn},
  \citenamefont {Grosnick}, \citenamefont {Kponou}, \citenamefont {Lee},
  \citenamefont {Li}, \citenamefont {Lopiano}, \citenamefont {Luccio},
  \citenamefont {Makdisi}, \citenamefont {Ratner}, \citenamefont {Reece},
  \citenamefont {Roser}, \citenamefont {Spinka}, \citenamefont {Ufimtsev},
  \citenamefont {Underwood}, \citenamefont {van Asselt}, \citenamefont
  {Williams},\ and\ \citenamefont {Yokosawa}}]{PhysRevLett.73.2982}%
  \BibitemOpen
  \bibfield  {author} {\bibinfo {author} {\bibfnamefont {H.}~\bibnamefont
  {Huang}}, \bibinfo {author} {\bibfnamefont {L.}~\bibnamefont {Ahrens}},
  \bibinfo {author} {\bibfnamefont {J.~G.}\ \bibnamefont {Alessi}}, \bibinfo
  {author} {\bibfnamefont {M.}~\bibnamefont {Beddo}}, \bibinfo {author}
  {\bibfnamefont {K.~A.}\ \bibnamefont {Brown}}, \bibinfo {author}
  {\bibfnamefont {G.}~\bibnamefont {Bunce}}, \bibinfo {author} {\bibfnamefont
  {D.~D.}\ \bibnamefont {Caussyn}}, \bibinfo {author} {\bibfnamefont
  {D.}~\bibnamefont {Grosnick}}, \bibinfo {author} {\bibfnamefont {A.~E.}\
  \bibnamefont {Kponou}}, \bibinfo {author} {\bibfnamefont {S.~Y.}\
  \bibnamefont {Lee}}, \bibinfo {author} {\bibfnamefont {D.}~\bibnamefont
  {Li}}, \bibinfo {author} {\bibfnamefont {D.}~\bibnamefont {Lopiano}},
  \bibinfo {author} {\bibfnamefont {A.~U.}\ \bibnamefont {Luccio}}, \bibinfo
  {author} {\bibfnamefont {Y.~I.}\ \bibnamefont {Makdisi}}, \bibinfo {author}
  {\bibfnamefont {L.}~\bibnamefont {Ratner}}, \bibinfo {author} {\bibfnamefont
  {K.}~\bibnamefont {Reece}}, \bibinfo {author} {\bibfnamefont
  {T.}~\bibnamefont {Roser}}, \bibinfo {author} {\bibfnamefont
  {H.}~\bibnamefont {Spinka}}, \bibinfo {author} {\bibfnamefont {A.~G.}\
  \bibnamefont {Ufimtsev}}, \bibinfo {author} {\bibfnamefont {D.~G.}\
  \bibnamefont {Underwood}}, \bibinfo {author} {\bibfnamefont {W.}~\bibnamefont
  {van Asselt}}, \bibinfo {author} {\bibfnamefont {N.~W.}\ \bibnamefont
  {Williams}}, \ and\ \bibinfo {author} {\bibfnamefont {A.}~\bibnamefont
  {Yokosawa}},\ }\href {\doibase 10.1103/PhysRevLett.73.2982} {\bibfield
  {journal} {\bibinfo  {journal} {Phys. Rev. Lett.}\ }\textbf {\bibinfo
  {volume} {73}},\ \bibinfo {pages} {2982} (\bibinfo {year}
  {1994})}\BibitemShut {NoStop}%
\bibitem [{\citenamefont {Barkov}\ \emph {et~al.}(1987)\citenamefont {Barkov}
  \emph {et~al.}}]{MRD_OMEGA_ENG}%
  \BibitemOpen
  \bibfield  {author} {\bibinfo {author} {\bibfnamefont {L.}~\bibnamefont
  {Barkov}} \emph {et~al.},\ }\href@noop {} {\bibfield  {journal} {\bibinfo
  {journal} {{JETP Let.}}\ }\textbf {\bibinfo {volume} {46}},\ \bibinfo {pages}
  {164} (\bibinfo {year} {1987})}\BibitemShut {NoStop}%
\bibitem [{\citenamefont {Nakamura}\ \emph {et~al.}(1998)\citenamefont
  {Nakamura}, \citenamefont {Drachenfels}, \citenamefont {Durek}, \citenamefont
  {Frommberger}, \citenamefont {Hoffmann}, \citenamefont {Husmann},
  \citenamefont {Kiel}, \citenamefont {Klein}, \citenamefont {Klein},
  \citenamefont {Menze}, \citenamefont {Michel}, \citenamefont {Nakanishi},
  \citenamefont {Naumann}, \citenamefont {Okumi}, \citenamefont {Reichelt},
  \citenamefont {Sato}, \citenamefont {Schoch}, \citenamefont {Steier},
  \citenamefont {Togawa}, \citenamefont {Toyama}, \citenamefont {Voigt},\ and\
  \citenamefont {Westermann}}]{NAKAMURA199893}%
  \BibitemOpen
  \bibfield  {author} {\bibinfo {author} {\bibfnamefont {S.}~\bibnamefont
  {Nakamura}}, \bibinfo {author} {\bibfnamefont {W.}~\bibnamefont
  {Drachenfels}}, \bibinfo {author} {\bibfnamefont {D.}~\bibnamefont {Durek}},
  \bibinfo {author} {\bibfnamefont {F.}~\bibnamefont {Frommberger}}, \bibinfo
  {author} {\bibfnamefont {M.}~\bibnamefont {Hoffmann}}, \bibinfo {author}
  {\bibfnamefont {D.}~\bibnamefont {Husmann}}, \bibinfo {author} {\bibfnamefont
  {B.}~\bibnamefont {Kiel}}, \bibinfo {author} {\bibfnamefont {F.}~\bibnamefont
  {Klein}}, \bibinfo {author} {\bibfnamefont {F.}~\bibnamefont {Klein}},
  \bibinfo {author} {\bibfnamefont {D.}~\bibnamefont {Menze}}, \bibinfo
  {author} {\bibfnamefont {T.}~\bibnamefont {Michel}}, \bibinfo {author}
  {\bibfnamefont {T.}~\bibnamefont {Nakanishi}}, \bibinfo {author}
  {\bibfnamefont {J.}~\bibnamefont {Naumann}}, \bibinfo {author} {\bibfnamefont
  {S.}~\bibnamefont {Okumi}}, \bibinfo {author} {\bibfnamefont
  {T.}~\bibnamefont {Reichelt}}, \bibinfo {author} {\bibfnamefont
  {H.}~\bibnamefont {Sato}}, \bibinfo {author} {\bibfnamefont {B.}~\bibnamefont
  {Schoch}}, \bibinfo {author} {\bibfnamefont {C.}~\bibnamefont {Steier}},
  \bibinfo {author} {\bibfnamefont {K.}~\bibnamefont {Togawa}}, \bibinfo
  {author} {\bibfnamefont {T.}~\bibnamefont {Toyama}}, \bibinfo {author}
  {\bibfnamefont {S.}~\bibnamefont {Voigt}}, \ and\ \bibinfo {author}
  {\bibfnamefont {M.}~\bibnamefont {Westermann}},\ }\href {\doibase
  https://doi.org/10.1016/S0168-9002(98)00298-8} {\bibfield  {journal}
  {\bibinfo  {journal} {Nuclear Instruments and Methods in Physics Research
  Section A: Accelerators, Spectrometers, Detectors and Associated Equipment}\
  }\textbf {\bibinfo {volume} {411}},\ \bibinfo {pages} {93 } (\bibinfo {year}
  {1998})}\BibitemShut {NoStop}%
\end{thebibliography}%
\end{document}